\documentclass[prb,aps,twocolumn]{revtex4}

\usepackage{graphicx}
\usepackage{amsmath}
\usepackage{epstopdf}
\usepackage{amsbsy}
\usepackage{appendix}
\usepackage{color}
\usepackage{dcolumn}% Align table columns on decimal point
\usepackage{bm}% bold math

\newcommand{\mb}{\mathbf}

\newcommand{\vk}{\mathbf{k}}

\newcommand{\vp}{\mathbf{p}}

\newcommand{\vrr}{\mathbf{r}}

\newcommand{\vx}{\mathbf{x}}

\newcommand{\va}{\mathbf{a}}
\newcommand{\vA}{\mathbf{A}}

\newcommand{\beq}{\begin{equation}}
\newcommand{\eeq}{\end{equation}}
\newcommand{\bea}{\begin{eqnarray}}
\newcommand{\eea}{\end{eqnarray}}

\begin{document}

\title{Is the composite fermion state of Graphene a doped Chern insulator?}
\author{Saurabh Maiti$^{1,2}$ and Tigran A. Sedrakyan$^1$}
\affiliation {$^1$Department of Physics, University of Massachusetts,
Amherst, Amherst, MA 01003}
\affiliation {$^2$Department of Physics, Concordia University, Montreal, QC H4B1R6}

\date{\today}

\begin{abstract}

Graphene in the presence of a strong external magnetic field is a unique attraction for investigations of the fractional quantum Hall (fQH) states with odd and even denominators of the fraction. Most of the
attempts to understand Graphene in the strong-field regime were made through exploiting the universal low-energy effective description of Dirac fermions emerging from the nearest-neighbor hopping model of electrons on a honeycomb lattice. We highlight that accounting for the next-nearest-neighbor hopping terms in doped Graphene can lead to a unique redistribution of magnetic fluxes within the unit cell of the lattice. While this affects all the fQH states, it has a striking effect at a half-filled Landau-level state: it leads to a composite fermion state that is equivalent to the doped
topological Chern insulator on a honeycomb lattice. At energies comparable to the Fermi energy, this state possesses a Haldane gap in the bulk proportional to the next-nearest-neighbor hopping and density of dopants. {We argue that this microscopically derived energy gap survives the projection to the lowest band. We also conjecture that the gap should be present in a microscopic theory giving the recently proposed particle-hole symmetric Dirac composite fermion scenario of the half-filled Landau-level. The proposed gap is lower than the chemical potential, and is predicted to be parametrically separated from the Dirac point in the latter description.} Finally we conclude by proposing experiments to detect this gap; the associated boundary mode; and encourage cold-atom setups to test other predictions of the theory.
\end{abstract}

\maketitle

\section{Introduction}

Since the discovery of the ``$1/3$-plateau"
in resistivity in the 2DEG Al-Ga-As\cite{Tsui1983}, the fractional
quantum Hall (fQH) effect has since intrigued the researchers. A
multitude of states between filling fractions $\nu=0$ and $\nu=1$
of the lowest Landau level has been seen in
2DEGs\cite{Kukushkin1999,Verdene2007} and Dirac
systems\cite{Dean2011,Du2009,Bolotin2009,Ghahari2011,Feldman2012,Feldman2013} with patterns consistent with $n/(pn\pm1)$ (the Jain
series\cite{Jain1989}; where $p$ is an even integer and
$n\in\{1,2,...\}$). The even denominator states are even more
exotic displaying anyonic behavior\cite{Zibrov2017}.

Besides the experimental progress, there have been significant contributions from the theoretical side. Major advancements were provided by the
formulation of the variational wavefunction by
Laughlin\cite{Laughlin1983}, the composite fermion theory by
Jain\cite{Jain1989,Jain2015}, the Chern-Simon's (CS) gauge theory
of the fQHE\cite{Zhang1989,Lopez1991,Lopez1995}, topological order
in fQHE states\cite{Wen1995}, effect of spin degree of
freedom\cite{Davenport2012}, to name a few. The physics of even
denominators were explored in the famous Halperin-Lee-Read (HLR)
theory of the half-filled Landau level\cite{Halperin1993}; the
Dirac composite fermion\cite{Son2015}; new field
theories\cite{Goldman2018}; and the vortex metal
state\cite{You2018}. The half-filled Landau level has itself been
an intense subject of
investigation\cite{Sedrakyan2008,Sedrakyan2009,Geraedts2016,Wang2017X}.

{ The composite fermion description of the fQH within HLR theory
has been one of the most productive ones. Recently, it has been
argued, however, that the HLR theory needs to be modified in order
to capture the correct symmetry properties (particularly the
particle-hole symmetry) of the low-energy effective theory
describing the $\nu=1/2$ state. Motivated by these new advances, a
question emerges whether the half-filled lowest Landau-level state
exhibits universal properties irrespective the microscopic
characteristics of the system and whether it represents a quantum
criticality and a stable phase of matter with well defined
low-energy quasiparticles. Experimentally it has been observed
that when the amount of the disorder is small, the Hall
resistivity exhibits a linear behavior near $\nu=1/2$ state as a
function of the magnetic field, $B$, between Jain's fractional
quantum Hall series. This observation, together with the fact that
Ohm resistivity is also finite and continuous,  does guarantee
stability of the compressible state. Although the theoretical
microscopic derivation of the state with correct symmetry
properties is still lacking, we side with the idea that there
exists modification of the flux attachment within the HLR
approach, that yields Son's Dirac composite fermions. Basing on
this idea, we will outline here novel properties of composite
fermions which are independent of a particular form of the flux
attachment. Hence we will apply HLR theory to describe our
predictions, conjecturing that the same universal features will
show up in any macroscopic derivation of Dirac composite fermions,
which is not only confined to the lowest energy sector.

The HLR theory is a continuous microscopic theory based on formally exact flux attachment procedure, where the composite fermion quasiparticle is obtained upon attaching an even number magnetic flux quanta to an electron.  To describe the $\nu=1/2$ state, one attaches just two fluxes followed by the flux smearing mean-field ansatz. In this way, the background magnetic field is ultimately canceled out yielding the HLR composite fermion sea as a ``variational" ground state for the system with Chern-Simons gauge fluctuations around it. }

\begin{figure}[tp]
$\begin{array}{cc}
\includegraphics[width=0.5\linewidth]{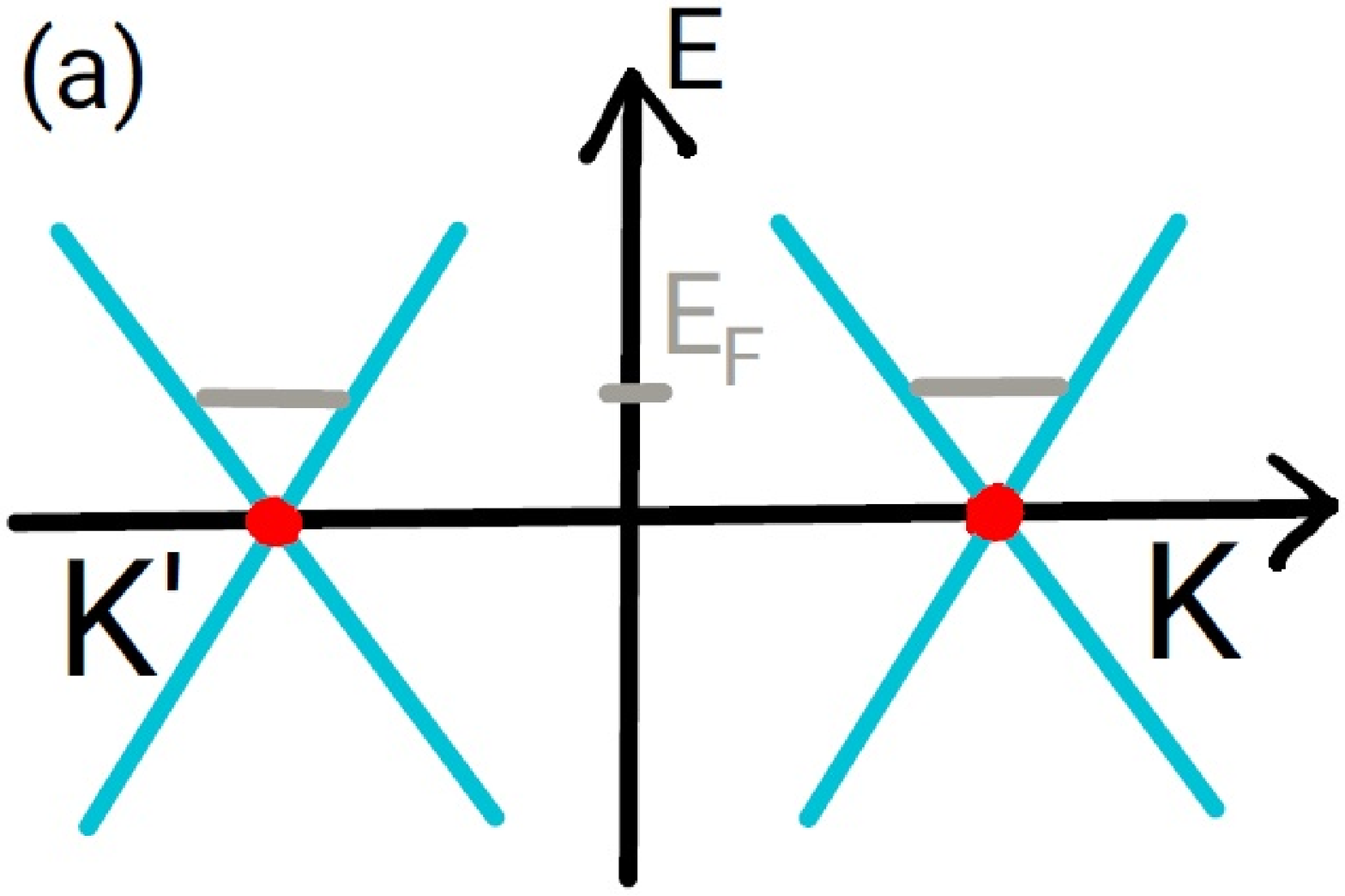}&
\includegraphics[width=0.5\linewidth]{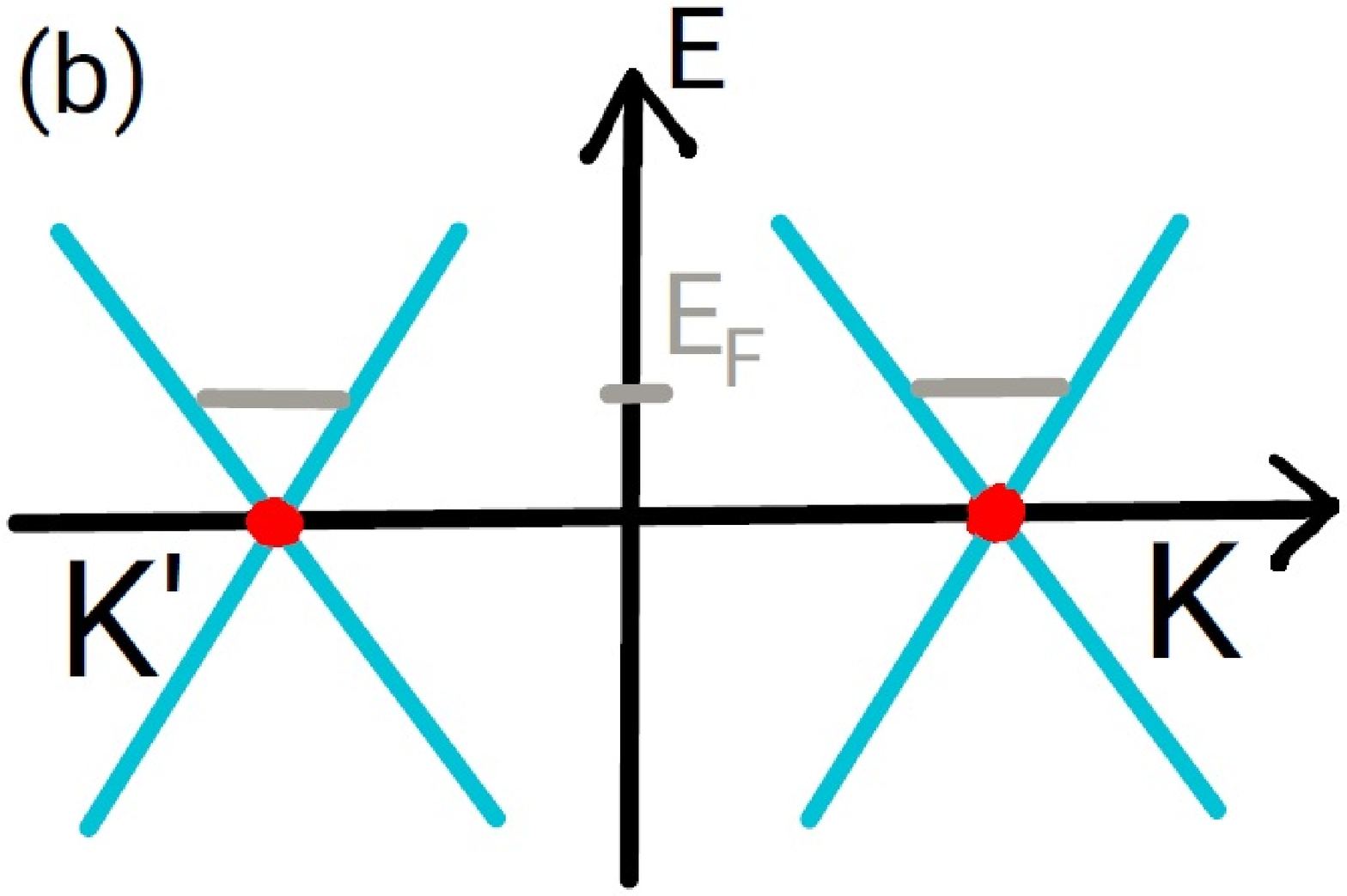}\\
\includegraphics[width=0.5\linewidth]{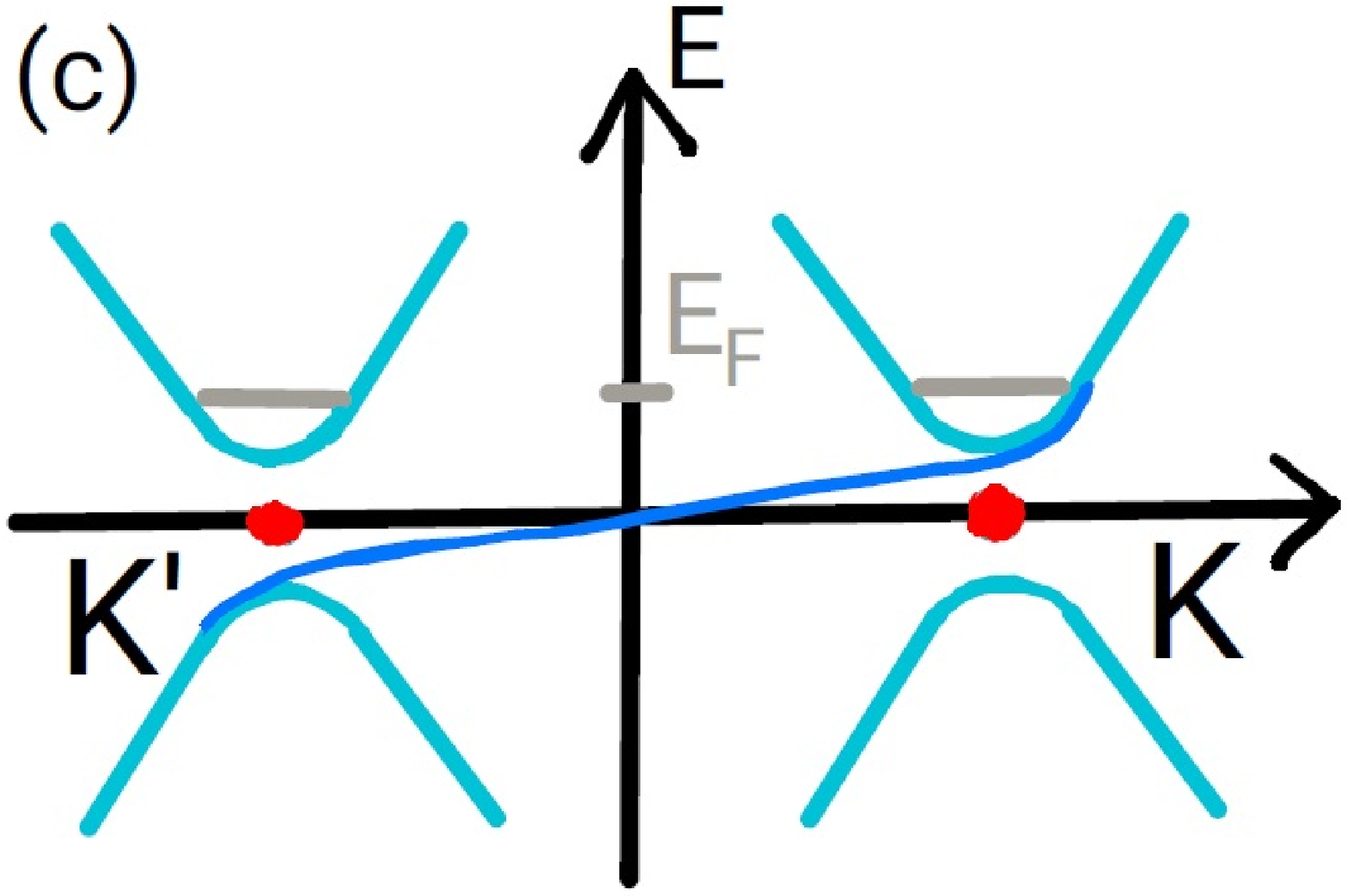}&
\includegraphics[width=0.5\linewidth]{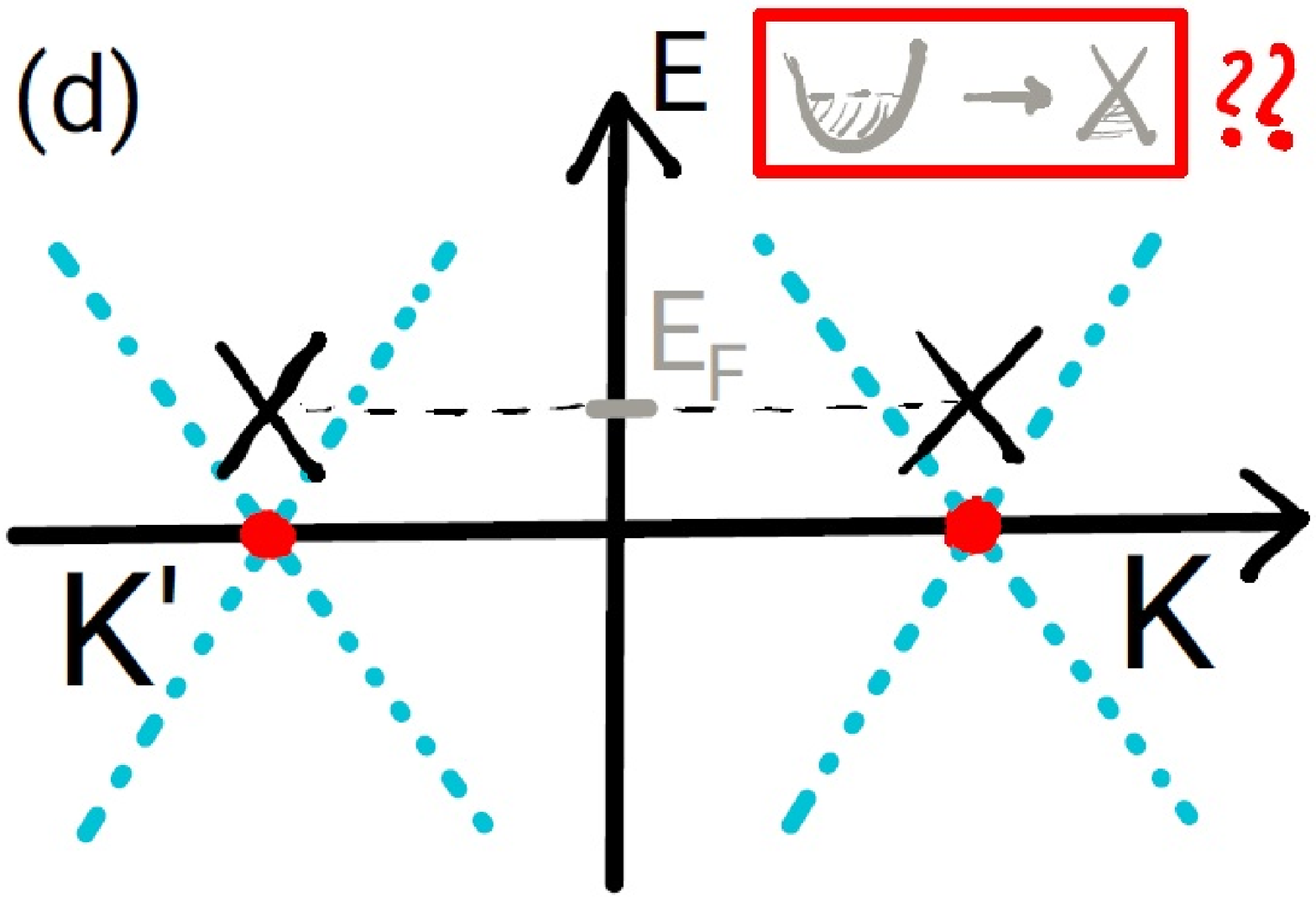}
\end{array}$
\caption{\label{fig:Schematic} (a) A schematic of the dispersion of electronic states in Graphene in zero field. (b) The spectrum of composite fermion states at half-filling condition based on naive application of HLR-type theory (uniform background field cancellation). (c) The composite fermion spectra based on this work. It consists of a bulk gap and an edge state. (d) Conjecture for how the Dirac composite fermion analog might look like. This is not addressed in this work. The inset represents the scenario for a parabolic band. $E_F$ is the Fermi level.}
\end{figure}

{ When extending this logic to lattices one introduces a lattice
Chern Simons theory and perform a mean field consistent with Gauss
law constraints. One can then remain in the lattice
picture\cite{Kumar2014,Sedrakyan2012,Sedrakyan2014,Sedrakyan2015,Maiti2018}
or take leap and consider the continuum limit after performing the
mean-field as in
Refs.~\onlinecite{Sedrakyan2012,Sedrakyan2014,Sedrakyan2015,Maiti2018}.
This latter approach is what we pursue in this work with the
intention to arrive at an HLR-like theory after the mean-field. We
explore what a CS flux distribution in a lattice may look like
(with nearest and next-neighbor hoppings) and the corresponding
continuum limit spectrum.} While studies have looked into the
effect of degeneracies on the fQH states in
lattices\cite{Sodeman2014,Wu2015}, some important considerations
regarding the role of next nearest neighbor (nnn) hoppings and the
flux distribution within a unit cell had been postponed to a
future consideration. Here we make an attempt to account for these
aspects in doped Graphene, and report a modification to the known
results especially at half-filling of the lowest Landau level. Our
first result is that at half-filling { of the Landau level, we do}
expect composite fermions. { However, one must contrast this
inference from those made with naive expectations. One could argue
that if we applied the HLR-idea to Graphene, the flux-detachment
from the composite fermions would lead to a complete cancellation
of the background magnetic field and thus the spectrum on the
composite fermions should be identical to doped Graphene in zero
field as shown in Fig. \ref{fig:Schematic}b. Our theory suggests
that while the low energy excitations may seem similar to the
naive HLR version, the spectrum of composite fermions has a
non-trivial Haldane type gap\cite{Haldane1988} in the bulk
spectra, with a chiral edge mode around the sample, as shown in
Fig. \ref{fig:Schematic}c. The essential physical property that
distinguishes the naive application of HLR from out theory is the
presence of the nnn hopping in Graphene (which is not
negligible\cite{CastroNeto2009,Kundu2011}). We emphasize that this
is not the Dirac-composite fermion theory\cite{Son2015} where the
postulate is that it is the correct theory of composite fermions
formed in a system of electrons in the continuum model with
parabolic dispersion.}

The second result we demonstrate is that in a generic fQH state there is a flux
distribution within a unit cell arising from the external and the emergent
CS fields. The former is uniformly distributed across the unit
cell and we refer to this type of distribution as the Maxwell
type. The latter results in a flux modulation within the unit cell which can be seen as a superposition of a Maxwell
type component $\phi_{MW}$ (which contributes to the total flux
through the unit cell), and an intra-unit cell component $\phi_H$
(which is only responsible for flux modulation and does not
contribute to any net flux through the unit cell). The flux
$\phi_{MW}$ is an orbital component that can be accounted for by
minimal coupling of a gauge field (that is given by the 2+1D CS
action), where as $\phi_H$ is an intra-unit cell property that
should be accounted for independently in the resulting
two-component Hamiltonian (the closest analogy would be the spin-field
coupling of Dirac electrons in the relativistic theory of electron).

In what follows, we will derive the origin of $\phi_H$ and argue
for its existence whenever there are internal hoppings within the unit
cell. We will apply this general scheme to nnn Graphene at half
filling and derive the effective low energy topological
Hamiltonian for composite fermions and present the flux
distributions for other filling fractions. We provide predictions
for the gap in terms of the lattice parameters of Graphene and external field and motivate some cold atom setups to test our theory.

\section{Peierls' substitution with nnn Graphene}

The first
intriguing observation is about the tight binding Hamiltonian for
Graphene with nnn hoppings. This can be written in the momentum
space as \bea\label{eq:HamG1} H&=&-t\left(
\begin{array}{cc}
rT^{A}_{\vk}&T_{\vk}\\
T^*_{\vk}&rT^{B}_{\vk}
\end{array}\right).
\eea Here, $t$ and $rt$ denote the nn and nnn hopping amplitudes
respectively, and \bea\label{eq:E}
T^{A}_{\vk}=T^{B}_{\vk}&=&e^{i\vk\cdot \va_1}+e^{i\vk\cdot
\va_2}+e^{i\vk\cdot \va_3}~+~{\rm c.c.},\nonumber\\
T_{\vk}&=&e^{i\vk\cdot \mb{e}_1}+e^{i\vk\cdot
\mb{e}_2}+e^{i\vk\cdot \mb{e}_3},\eea where
$\mb{e}_1=\frac{a}{\sqrt{3}}(\frac{\sqrt{3}}{2},\frac{-1}{2});
~\mb{e}_2=\frac{a}{\sqrt{3}}(0,1),
~\mb{e}_3=-\frac{a}{\sqrt{3}}(\frac{\sqrt{3}}{2},\frac{1}{2})$
(the translations from A to B atoms of Graphene), and
$\va_1=a(1,0),~\va_2=a(\frac{1}{2},\frac{\sqrt{3}}{2}),
~\va_3=a(\frac{1}{2},\frac{-\sqrt{3}}{2})$ (the lattice
translation vectors), where $a$ is the lattice constant. Observe
that $T^{A}_{\vk}=T_{\vk}T^*_{\vk}-3$ and
$T^{B}_{\vk}=T^*_{\vk}T_{\vk}-3$. When an external field $\mb{B}$
is applied, one employs the Peierls' substitution and either
resorts to a Hofstadter like scheme\cite{Hofstadter1976} to obtain
the spectrum, or performs an expansion of $H$ around a high
symmetry point ($\vk_0$) in the Brilliouin zone(BZ) where the
chemical potential is expected to lie, in powers of
$\delta\vk=\vk-\vk_0$ and setting $\delta\vk\rightarrow
-i\partial_{\vrr}+e\vA\equiv\vp$, such that
$\mb{\nabla}\times\mb{A}=\mb{B}$.
%($-e$ is the electronic charge
%that couples the electronic system to the gauge field $\mb{A}$).
We shall refer to this procedure as elongating the momentum $\vk$.

\begin{figure}[htp]
\includegraphics[width=0.99\linewidth]{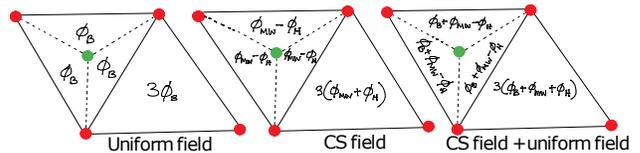}
\caption{\label{fig:Bflux}Flux distribution in a unit cell of
Graphene in uniform external field, CS field, external$+$CS field.
The CS field consists of a Maxwell part $\phi_{MW}$ and a
modulated part $\phi_H$ ($=\phi_{MW}$) that does not contribute to
the total flux. In the formulas we use the dimensionless fluxes $\phi_{b,h,mw}$ which are defined at appropriate places.}
\end{figure}

This seemingly simple prescription has led to many useful results
in many lattices including Graphene (with $r=0$). When $r\neq0$,
the first thing we note is that upon momentum elongation
$\vk\rightarrow\vk_0+\vp$, $T^{A}_{\vk_0+\vp}\neq
T_{\vk_0+\vp}T^*_{\vk_0+\vp}-3$. This is because,
\beq\label{eq:commRel} e^{i\vp\cdot\mb{e}_1}e^{-i\vp\cdot\mb{e}_2}
=e^{i\vp\cdot\va_3}e^{-i2\pi\phi_B/\phi_0},\eeq where
$\phi_B=Ba^2/4\sqrt{3}$, the flux through the small triangle in
Fig. \ref{fig:Bflux}; and $\phi_0=2\pi/e$ the flux quantum
($\hbar=1$). One is thus left with a choice of using either
$T^{A}_{\vk_0+\vp}$ or $T_{\vk_0+\vp}T^*_{\vk_0+\vp}-3$ in the
Hamiltonian.

This ambiguity is removed by noting that the two choices reflect
the two translations from $A\rightarrow A$: directly
($T^{A}_{\vk}$) or via $B$ ($T_{\vk}T^*_{\vk}$), which must not
commute as it encloses a flux $\phi_B$. The choice of
$T_{\vk_0+\vp}T^{*}_{\vk_0+\vp}$ is consistent with $\phi_B=0$ and
hence we can conclude that the correct choice is using
$T^{A/B}_{\vk_0+\vp}$. In other words, to obtain a Maxwell type
flux distribution in a lattice, the momentum elongation must be
carried out on the translations corresponding to every allowed hop
on the lattice. Carrying this procedure out and expanding the
Hamiltonian around the $K$-point of Graphene
[$\vk_0=\frac1a(\frac{4\pi}{3},0)$] to $\mathcal{O}(p^2)$, we find
\bea\label{eq:Heff} H_B(\vrr)&=&-t\left[r\bar p^2\sigma_0-\bar
p_x\sigma_x-\bar p_y\sigma_y\right],\eea where $\bar p_i\equiv
\sqrt{3}ap_i/2$. This effective low energy Hamiltonian can be
exactly solved. The energy spectrum is \bea\label{eq:spectrum}
\frac{E^{\pm}}{-t}&=&2r\phi_b n\pm\sqrt{2\phi_b n +
r^2\phi_b^2},\eea where $\phi_b\equiv3\sqrt{3}e\phi_B$. The
wavefunctions for the $A/B$ components for the quantum number $n$
with the gauge choice of $\vA=-By\hat\vx$ are:
$\psi^{\pm}_B=\Phi_n(\vrr)$,
$\psi^{\pm}_A=\Phi_{n-1}(\vrr)/(c\pm\sqrt{1+c^2})$, where
$c=r\sqrt{\phi_b/2n}$, and \beq\label{eq:nnWF} \Phi_n(\vrr)=
\frac{1}{\sqrt{2^n n!\sqrt{\pi}}}e^{ikx}e^{-\bar y^2/2}H_{n}(\bar
y).\eeq Here $\bar y\equiv (y-kl^2_B)/l_B$, $\l^2_B=1/Be$, $k$ is
a quantum number corresponding to translations along $x$, and
$n\in\{1,2,...\}$. For $n=0$, we have $-E/t=r\phi_b$ and the
wavefunction components are $\psi_B=\Phi_0(\vrr)$ and $\psi_A=0$.
At the $K'$ point where
$\vk_0=\frac1a\left(\frac{-4\pi}{3},0\right)$, our low energy
Hamiltonian is the same as $H_B$ but with $p_x\rightarrow-p_x$.
This leads to the same spectrum however,
$\psi^{K}_B\rightarrow\psi^{K'}_A$ and
$\psi^{K}_A\rightarrow-\psi^{K'}_B$. Thus the $n=0$ landau level
at $K'$ point only has sublattice $A$ occupied. The presence of
nnn affects the relative weights of the respective sublattices. It
must be noted that the expansion to $\mathcal{O}(p^2)$ is only
valid for weak fields such that $\phi_B/\phi_0\ll1$.

At finite $r$, there is an interesting situation that arises
specifically when $\phi_B=\phi_0/6$ (see Fig. \ref{fig:Bflux}).
The total flux through the unit cell is $\phi_0$ meaning the phase
accumulation around the unit cell is $2\pi$ which restores the
translation invariance to the lattice, although the internal
hoppings still acquire a Peierls' phase. The distribution of this
phase is demonstrated in the appendix. The
effective Hamiltonian (at $K$ point) is \bea\label{eq:Heffff}
H_B(\vrr)&=&-t\left[r\bar k^2\sigma_0-\bar k_x\sigma_x-\bar
k_y\sigma_y+r\phi_b\sigma_z\right].\eea The spectrum is
$-E^{\pm}/t=r\bar k^2\pm\sqrt{\bar k^2+r^2\phi_b^2}$, with a
spectral gap of $2r\phi_b|t|$. At such a value of the external
field the system becomes equivalent to the Haldane's model for
Graphene\cite{Haldane1988}. Given the lattice constant of Graphene
of $a=2.46\AA$, this phenomenon is expected to happen when
$6\phi_B=\frac{\sqrt{3}}{2}Ba^2=\phi_0$, resulting in an
unrealistic magnetic field of $B\approx8\times10^4$T. Below we
will show that the Haldane's model can be realized for composite
fermions at much weaker fields. But before we make this
connection, we describe how the CS field affects the lattice and
the flux distribution within the unit cell.

\section{The CS field in nnn Graphene}

Consider a system of N
particles with sets of coordinates
$\{\vrr\}\equiv\{\vrr_1,...,\vrr_N\}$. The many-body state
$\Psi(\{\vrr\})$ constructed as a Slater determinant of the single
particle states obeys $H_{\rm
kin}(\{\vrr\})\Psi(\{\vrr\})=E\Psi(\{\vrr\})$ (where `kin' denotes
the kinetic part of the Hamiltonian). If the single particle
states are picked from a manifold of degenerate states, one can
propose another solution, without costing any energy, by
introducing composite fermions in terms of original electrons
coupled via a CS phase \bea\label{eq:CS}
\Psi(\{\vrr\})=e^{i\Lambda_{\{\vrr\}}}\Psi_c(\{\vrr\}),~
\Lambda_{\{\vrr\}}=\kappa\sum_{\vrr'\neq\vrr}\theta_{\vrr'\vrr},&&
\eea where $\theta_{\vrr'\vrr}=\arg (\vrr'-\vrr)$, and $\kappa$ is
an even integer to retain fermionic statistics of the resulting
composite particles. It then follows that: \bea\label{eq:MBG}
&H(\{\vrr\})\Psi(\{\vrr\})&=E\Psi(\{\vrr\}),\nonumber\\
\Rightarrow &H(\{\vrr\},\vA_{\{\vrr\}})\Psi_c(\{\vrr\})&=E\Psi_c(\{\vrr\}),\nonumber\\
{\rm MFA\Rightarrow}&\left[\sum_{i}H_{\rm
kin}(\vrr_i,\vA^{MW}_{\vrr_i})\right]\Psi^{\rm
sl}_c(\{\vrr\})&=E\Psi^{\rm sl}_c(\{\vrr\}),\eea where
$\vA_{\{\vrr\}}(\vrr_i)=\frac1e\partial_{\vrr_i}\Lambda_{\{\vrr\}}$;
$H(\{\vrr\},\vA_{\{\vrr\}})
=e^{-i\Lambda_{\{\vrr\}}}H(\{\vrr\})e^{i\Lambda_{\{\vrr\}}}$; and
${\rm sl}$ stands for Slater determinant. The standard
flux-smearing mean-field approximation\cite{Halperin1993} (MFA)
designed to capture the transition to a topologically non-trivial
system $H(\{\vrr\},\vA_{\{\vrr\}})\rightarrow
\sum_{\{\vrr\}}H_{\rm kin}(\vrr,\vA^{MW}_{\vrr})$. This amounts to
changing the non-local $\vA_{\{\vrr\}}$ to a local single-particle
$\vA^{MW}(\vrr)$. Equivalently, the CS magnetic field defined by
$\mb{B}(\vrr)\equiv\nabla_{\vrr}\times\vA_{\{\vrr\}}
=\phi_0\kappa\sum_{\vrr'\neq\vrr}\delta(\vrr'-\vrr)$ is
approximated by a uniform
$\mb{B}=\phi_0\kappa\rho_{2D}=\nabla_{\vrr}\times\vA^{MW}_{\vrr}$.
This is the many-body version of the field theories considered in
Refs. \cite{Lopez1991,Halperin1993}. This formalism has been used
in Refs. \cite{Sedrakyan2012,Sedrakyan2014,
Sedrakyan2015,Sedrakyan2015_2,Maiti2018} to treat hard-core bosons
as fermions with an odd $\kappa$ leading to chiral spin-liquid
behaviour in honeycomb and Kagome lattices.

As discussed in Refs. \cite{Sedrakyan2014,Maiti2018}, the CS field
produces a flux that is bound to a particle, the flux enclosed
within the space of the particles in a unit cell must be zero. The
net flux that arises from the MFA must then be re-distributed to
the part of the unit cell that does not include any internal hoppings.
Thus, the regions bounded by internal hoppings that are entirely
within the unit cell (defined as loops having at the most one
shared edge with external cells) must enclose zero flux. This is a
constraint in our MFA to account for the CS character and
distinguish it from a Maxwellian field. This constraint is
efficiently implemented by introducing two types of fluxes:
$\phi_{MW}$ (which contributes to the total flux per unit cell as
constrained by the MFA) and $\phi_H$ (which does not contribute to
the flux per unit cell). Our constraint requires
$\phi_H=\phi_{MW}$ (see Fig. \ref{fig:Bflux}).

Note that this constraint implies that the direct
$A\rightarrow A$ translation and the one mediated through a $B$
atom commute (which is different from the case of uniform field).
Also note that while $\phi_{MW}$ can be accounted for by momentum
elongation, $\phi_H$ cannot as it emerges as an internal degree of
freedom due to the CS constraint. This results in the effective
Hamiltonian (at the $K$ point) to be \bea\label{eq:HeffCS}
H_{CS}(\vrr)&=&-t\left[r\bar p^2\sigma_0-\bar p_x\sigma_x-\bar
p_y\sigma_y+r\phi_h\sigma_z\right],\eea where $\bar
\vp=\sqrt{3}a[-i\partial_{\vrr}+e\vA^{MW}(\vrr)]/2$, and
$\phi_h=3\sqrt{3}e\phi_H$. Further, $[\bar p_x,\bar
p_y]=3\sqrt{3}\phi_{MW}$. Since the case of CS requires
$\phi_{MW}=\phi_H$, it is possible to compact Eq.
(\ref{eq:HeffCS}): \bea\label{eq:HeffCSX}
H_{CS}(\vrr)&=&-t\left[r\left\{\bar
\vp\cdot\mb{\sigma}\right\}^2-\bar \vp\cdot\mb{\sigma}\right],\eea
where the dot-product is 2-dimensional. Thus,
$H_{CS}=-rt[H_{\vk_0+\vp}]^2-tH_{\vk_0+\vp}$. The CS nature of the
field relates flux to density and requires that
$\phi_{MW}=2\pi\kappa n_{uc}\nu_l/e$, where $n_{uc}$ is the number
of atoms per unit cell and $\nu_l$ is the deviation from half-filling
per site. The energy spectrum is: \bea\label{eq:HCSsol}
\frac{E^{\pm}}{-t}&=&2r|\phi_{mw}| n\pm\sqrt{2|\phi_{mw}| n}, \eea
where $\phi_{mw}=3\sqrt{3}e\phi_{MW}$, $n\in\{1,2,...\}$ and the
wavefunctions are $\psi^{\pm}_B=\Phi_n(\vrr)$, and
$\psi^{\pm}_A=\pm\Phi_{n-1}(\vrr)$. For $n=0$, $E=0$ and
$\psi_{B}=\Phi_0(\vrr)$ and $\psi_A=0$. At $K'$ point
$p_x\rightarrow-p_x$, $\phi_h\rightarrow-\phi_h$, and $\psi^{K}_B\rightarrow\psi^{K'}_A$,
$\psi^{K}_A\rightarrow-\psi^{K'}_B$.

\section{Composite Fermions in Graphene with nnn hoppings}

The degenerate
manifold that often motivates the use of the CS-phase attached
to the many-body wavefunction can be thought of as being provided
by an external field $\mb{B}$\cite{Halperin1993,Son2015}. We then
have to account for three types of fluxes within the unit cell:
$\phi_B$, $\phi_{MW}$ and $\phi_H$. As discussed, the first two produce a field that grows with area and can be accounted for by
momentum elongation, while $\phi_H$ needs to be introduced at the
level of matrix elements for the Hamiltonian (see appendix for detailed
construction). The resulting theory is a fermion model populating
Haldane's model coupled to the CS action that was generated from
flux attachment. This ``statistical" CS term is the same as the
one obtained in previous theories in the
literature\cite{Zhang1989,Naverich2001}. The distinguishing
feature however is the flux distribution within the unit cell
which is shown in Fig. \ref{fig:Bflux}. The resulting composite
fermion(CF) Hamiltonian around the $K$-point is:
\beq\label{eq:HeffCS2} H_{\rm CF}(\vrr)=-t\left[r\bar
p^2\sigma_0-\bar p_x\sigma_x-\bar
p_y\sigma_y+r\phi_{h}\sigma_z\right],\eeq where
$\vp=-i\partial_{\vrr}+e\vA^B_{\vrr}+e\vA^{MW}_{\vrr}$,
$\nabla_{\vrr}\times\vA^B=\mb{B}$, which leads to the flux
$\phi_B$, and $\nabla_{\vrr}\times\vA^{MW}=\mb{B}^{MW}$ which
leads to the flux $\phi_{MW}=\phi_H=\phi_0\kappa n_{uc}\nu_l$ from
the induced CS field. Here $\kappa=-2~{\rm sgn}(B)$, where
sgn($B$) simply indicates that the sign of $\kappa$ is such that
the induced field opposes the external field. Just like in the HLR
theory, the net `orbital' field (resulting from the vector
potential) experienced by the composite fermions is $\mb{B}_{\rm
eff}=\mb{B} + \mb{B}^{MW}=\phi_0\rho_{2D}(1/\nu+\kappa)\hat
B=\mb{B}(1-2\nu)$. But unlike the HLR theory, the two component
nature of the lattice causes the composite fermions to experience
an additional field that acts oppositely on the two atoms in the
unit cell. This effect is captured by the flux $\phi_H$ and is
analogous to the `spin' coupling of the Maxwell field to the true
spin-up/down fermions. The resulting spectrum is:
\beq\label{eq:HCSsol2}
\frac{E^{\pm}}{-t}=2r|\phi_{b}+\phi_{mw}|n\pm\sqrt{2|\phi_{b}+\phi_{mw}|
n+r^2\phi_b^2}, \eeq where $n\in\{1,2,...\}$, and for $n=0$ only
$E^{+}$ is present. The wavefunctions are similar to the solutions
for uniform field with $c\rightarrow\tilde
c=r\sqrt{\phi_b/2n|1+\phi_{mw}/\phi_b|}$. Note that Eq.
(\ref{eq:HeffCS2}) is different from Eq. (\ref{eq:HeffCS}) in the
definition of $\vp$. Due to the presence of the external field, we
cannot compact the Hamiltonian with a $(\mb{\sigma}\cdot\vp)^2$
term.

There are a number of points that need emphasis:(a) At
half-filling ($\nu=1/2$), we have $\mb{B}_{\rm eff}=0$ and
$\phi_{mw}=-\phi_b$. The spectrum of composite fermions is given
by $E^{\pm}$ listed after Eq. (\ref{eq:Heffff}) with the a
spectral gap of $r\phi_b|t|$. (b) The resulting flux distribution
in Fig. \ref{fig:Bflux} at $\nu=1/2$ (with $\phi_{mw}=-\phi_b$)
suggests that this is indeed the distribution considered by
Haldane\cite{Haldane1988}. (c) Just like the Landau-problem, Eq.
(\ref{eq:HCSsol2}) is to be seen as a solution to the composite
fermions in a weak $\mb{B}_{\rm eff}$, which is realized by small
departures from half-filling.

Noting that $\phi_B=\phi_0(n_{uc}\nu_l)/6\nu$ and
$\phi_{MW}=\phi_0(n_{uc}\nu_l)\kappa/6$, treating $\phi_B$ and
$\nu_l$ as external control variables, the $\nu=1/2$ condition is
realized whenever $\nu_l$ and $\mb{B}$ satisfy
$\phi_0\kappa(n_{uc}\nu_l)=\sqrt{3}Ba^2/2$. Departures from
$\nu=1/2$ can thus be achieved by slightly changing the field or
$\nu_l$ (which can be achieved by changing the chemical
potential). This mismatch creates the effective field that the
composite fermions are subjected to. In fact, for small deviations
such that $\phi_b=-\phi_{mw}+\delta$ ($\delta\ll\phi_b$), we get
$-E^{\pm}/t\approx \pm r\phi_b(1\pm2n|\delta|/\phi_b)$.
\begin{figure}[htp]
\includegraphics[width=0.99\columnwidth]{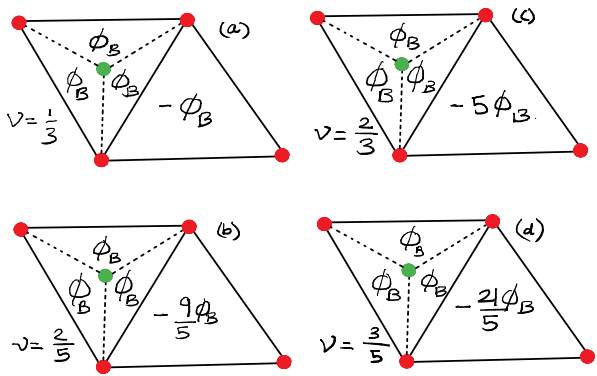}
\caption{\label{fig:fQHE}Flux distribution in a unit cell of
Graphene for various fQHE states given by the Jain series.}
\end{figure}

We thus observe that the necessary characteristic of the composite
fermion state in Graphene with nnn is the formation of the
Haldane's Chern insulator populated by composite fermions which,
in turn, are coupled to the fluctuating ``statistical" CS gauge
field with the coefficient $\propto 1/4\pi\kappa$. The low-energy
field theory description of this state is obtained upon
integration over fermionic degrees of freedom. This results in an
additional CS-like term, which combines with the statistical term
resulting in a coefficient { $\propto (1/4\pi\kappa+2
\text{sgn}(B)/8\pi)\neq0$  for the pure CS field. At this level,
the low-energy description of the composite fermion state of
Graphene does support an additional Chern-Simons action of the
fluctuating field, by which it differs from  the Dirac composite
fermion construction by Son\cite{Son2015} for a non-Dirac system.
However, the low energy quasiparticles are still Dirac fermions,
which is not surprising}, since the original dispersion of
electrons is not expected to qualitatively affect the ground state
properties at very low energies because of complete quenching of
the electronic kinetic energy by the strong magnetic
field\footnote{The connection with the Dirac composite fermion
construction by Son\cite{Son2015} for a non-Dirac system in a
magnetic field is set as a future goal and is beyond the scope of
this work}.

\begin{figure*}[htp]
$\begin{array}{cc}
\includegraphics[width=1\columnwidth]{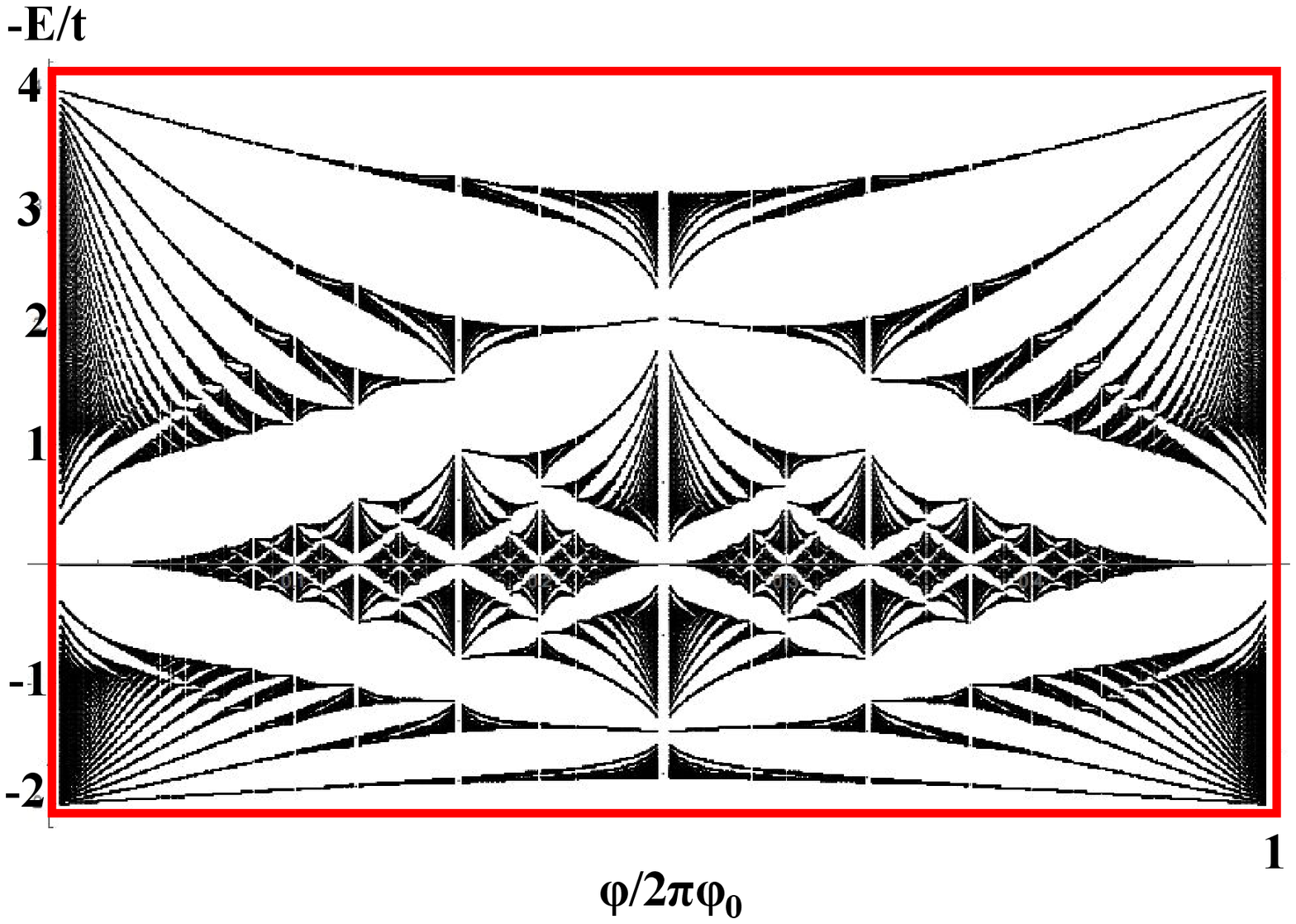}&
\includegraphics[width=1\columnwidth]{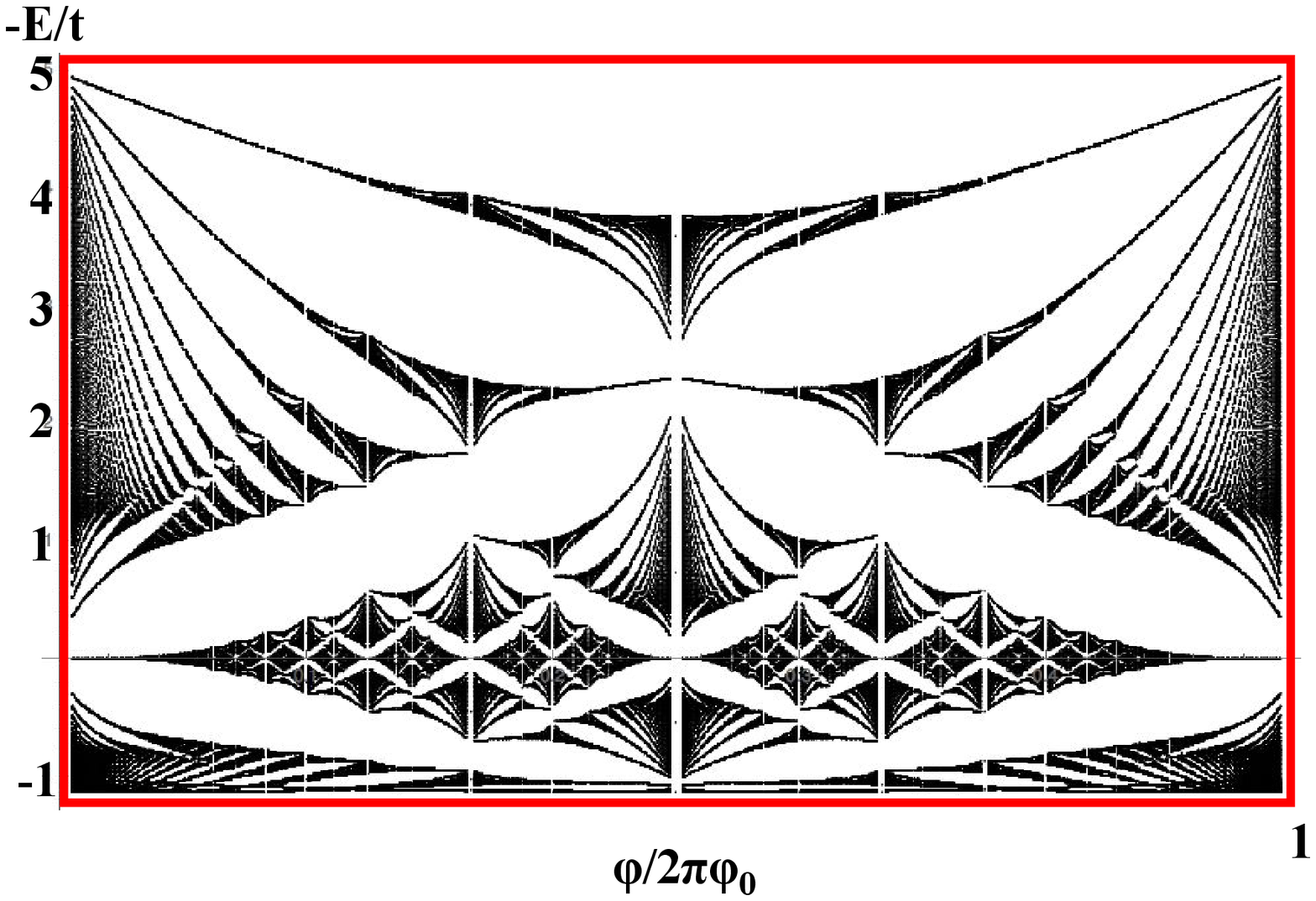}
\end{array}$
\caption{ \label{fig:Hoff}The Hofstadter spectrum for a CS field acting on nnn Graphene with $r=0.11$ (left) and $r=0.22$ (right). Observe the linear scaling and the flatness of the lowest energy level in the two plots.}
\end{figure*}
% In contrast, at $r=0$, integration over composite fermions yields a Maxwell
%term\cite{semenoff}, which, combined with the statistical CS,
%yields a Maxwell-CS theory\cite{dunne}.

This prescription also works for any other fQHE state. Consider
the Jain series: $\nu = n/(2n+1)$ and $\nu=(n+1)/(2n+1)$. { It is
easy to see that the total flux through the unit cell is given by
$6(\phi_B+\phi_{MW})$. However, as discussed earlier, this flux is
distributed as $\phi_B$ in each of the smaller triangles and
$3(\phi_B+2\phi_{MW}$) in the larger triangle. Further, recall
that $\phi_{MW}=\kappa\phi_0=-2\phi_0$, and $\phi_B=\phi_0/\nu$.
Expressing everything in terms of $\phi_B$, the flux that is
generated by the external magnetic field, the flux through the
smaller triangles is always $\phi_B$, though the larger triangle
is $3\phi_B(1-4\nu)$, and through the unit cell is
$6\phi_B(1-2\nu)$.} The resulting flux distributions for some of
the fQHE states in the Jain series are shown in Fig.
\ref{fig:fQHE}. The low energy Hamiltonian for small deviations
around each $n$ can be derived by appropriately enlarging the
unit-cell (as in the original Hofstadter
construction\cite{Hofstadter1976}): with the special care that
$\phi_H$ has to be accounted for every matrix element of the
Hamiltonian for the enlarged unit cell. {A quantitative account of
such flux modulation and its effect on the many-body wavefunction
remains an open question that will provide a valuable information
on Jain's states.}

\section{Experimental consequences}

Our theory suggests that in
addition to the expected Fermi-surface(FS), we must see a spectral
gap in the bulk consistent with the Haldane's model (Fig
\ref{fig:Schematic}b). Following the formulas in the text, for a
density of $\rho_{2D}=\bar\rho\times10^{11}$cm$^{-2}$, this gap is
$\approx r\bar\rho|t|$ meV (where $t$ must be in eV); and this
happens at a $B\approx8\bar\rho$ T. In Graphene, $t\approx 3$ eV
and $r\sim 0.1$\cite{CastroNeto2009}. If $\bar\rho$ is tuned from
$1-10$, then for $B\sim 8$-$80$T, we should see a Haldane gap of
$0.3$-$3$ meV. Further, $v\approx1\times10^6$ms$^{-1}$ implies
$E_F\approx8\sqrt{\bar\rho}$ meV which will range from
$8$-$24$meV. The energy scale suggests that the Haldane gap can be
probed by LDOS measurements via STM. Due to correspondence with
Haldane's model, the composite fermion bands are topological with
Chern number $\pm1$ and thus have chiral edge states which can be
measured via tunneling into the edge
\cite{PhysRevLett.80.141,PhysRevLett.102.096806,
PhysRevLett.114.156401} and via measuring the quantized thermal
transport due to a Luttinger liquid description of edge
states\cite{fisher,PhysRevB.98.155321}. The presence of a gapped spectrum for
$\nu=1/2$ state in Graphene is in line with findings in the two
component AlAs\cite{Zhu2018}.

{ \subsection{Implication for cold atoms} There is an interesting
consequence of our theory when applied to fermionic/bosonic cold
atoms at low densities. When the nnn hopping is present ($r\neq
0$), a characteristic feature at $r=1/6$ distinguishes the flux
distribution in our theory from those that might be predicted by
other theories: when $r>1/6$, the minimum of the lattice spectrum
in the presence of the external field (which forms from states
around the $\Gamma$-point of the original system), is
non-dispersive as shown in Fig. \ref{fig:Hoff}.

To understand this, consider the $\Gamma$-point the effective low energy Hamiltonian:
\beq\label{eq:HGamma}
H^{\Gamma}=-t\left(\begin{array}{cc}
r(9-2\bar p^2)&3-\frac{1}{3}\bar p^2\\
3-\frac{1}{3}\bar p^2&r(9-2\bar p^2)
\end{array}
\right).\eeq
The spectrum is given by
\beq\label{eq:Gammasol}
\frac{E^{\pm}}{-t}=9r\pm3-2\left(r\pm\frac16\right)(2n+1)\phi_b,
\eeq
and the wavefunctions are $\psi^{\pm}_A=\pm\psi^{\pm}_B=\Phi_n(\vrr)$, with $n\in\{0,1,2,...\}$. Thus for $r<1/6$, the energy of the lowest landau level, which can occur at any $n$, linearly increases with field. However, for $r>1/6$, the spectrum in Eq. (\ref{eq:Gammasol}) suggests that the energy goes down. This is indeed what happens for the uniform field case. However, for the CS case with $r>1/6$, the minimum of the energy for any $\{n,\phi_{mw}\}$ is actually locked at $E=-|t|/4r$. This means that the lowest energy level of the spectrum is dispersionless with the field $\phi_b$. This is not evident from Eq. (\ref{eq:Gammasol}) due to the fact that we expanded to $\mathcal{O}(p^2)$.

This minimum is guaranteed because the Hamiltonian is of the form $rG^2_{\vp}+G_{\vp}$. This means the eigenvectors of $G_{\vp}$ are also the eigenvectors of $G^2_{\vp}$, with squared eigenvalues. The quadratic form suggests that there is a minimum of the energy for any eigenvalue. For $r<1/6$, this minimum energy scales with the field as $E=(1/6-r)\phi_b$, but for $r>1/6$ the minimum is at $E=t/4r$ for infinite pairs of $\{n,\phi_{mw}\}$. Note that this effect of CS field also applies to the $K/K'$ point analysis. This we should expect a dispersionless band around the center of the spectrum. These cases are demonstrated in Fig. \ref{fig:Hoff}. In other words, if we were to tune $r$, our theory would predict that the lowest band in the plot for $E(\phi)$ will have a slope that decreases as $r$ increases and stall at zero slope (dispersionless). In other flux distributions, the slope becomes negative.}

\section{Concluding remarks}

To summarize, we showed that Graphene with nnn hoppings near
half-filling is described by a model of composite fermions living
on Haldane's Chern insulator model of Graphene. This perspective
not only provides an internal flux distributions within the unit
cell but also the effective wavefunctions for fQH states. {Very
interestingly, this perspective when applied to fermionization of
bosons in flat bands (as done in Ref. \onlinecite{Maiti2018}),
leads to chiral-spin liquid behavior. Thus besides motivating
experiments, this work potentially provides a unified picture of
treating chiral spin-liquids in spin-systems and fQH states in
fermionic systems. We believe that this motivates further research
and exploration of other fractional Quantum Hall systems along
similar lines to compare and contrast with existing literature and
guide future experiments}.

{ At this stage one may ask the question: why are we doing an HLR
type flux attachment instead of building a Dirac composite
fermion\cite{Son2015} analog? While we don't directly answer this
question, we point out that both HLR theory and the Dirac
composite fermion theories ultimately give the same result for the
observables. Since the bulk gap predicted in out theory is in
principle an observable, we expect this result to be robust. The
main reason we did not attempt the Dirac composite fermion analog
is that it is necessarily limited to the low energy excitations
and does not immediately address the presence of such a bulk gap.
However, the topological properties of the chiral bands must be
reproduced is such an approach and is an interesting question to
address next.}

%\textit{Acknowledgments}:
\section*{Acknowledgement}
We would like to thank A. Kamenev for
discussions. The research was supported by startup funds from
UMass Amherst.

%\begin{appendices}
\appendix

\section{Peierls phases for uniform and Chern Simons' fields}

\begin{figure*}[htp]
$\begin{array}{c}
\includegraphics[width=0.8\linewidth]{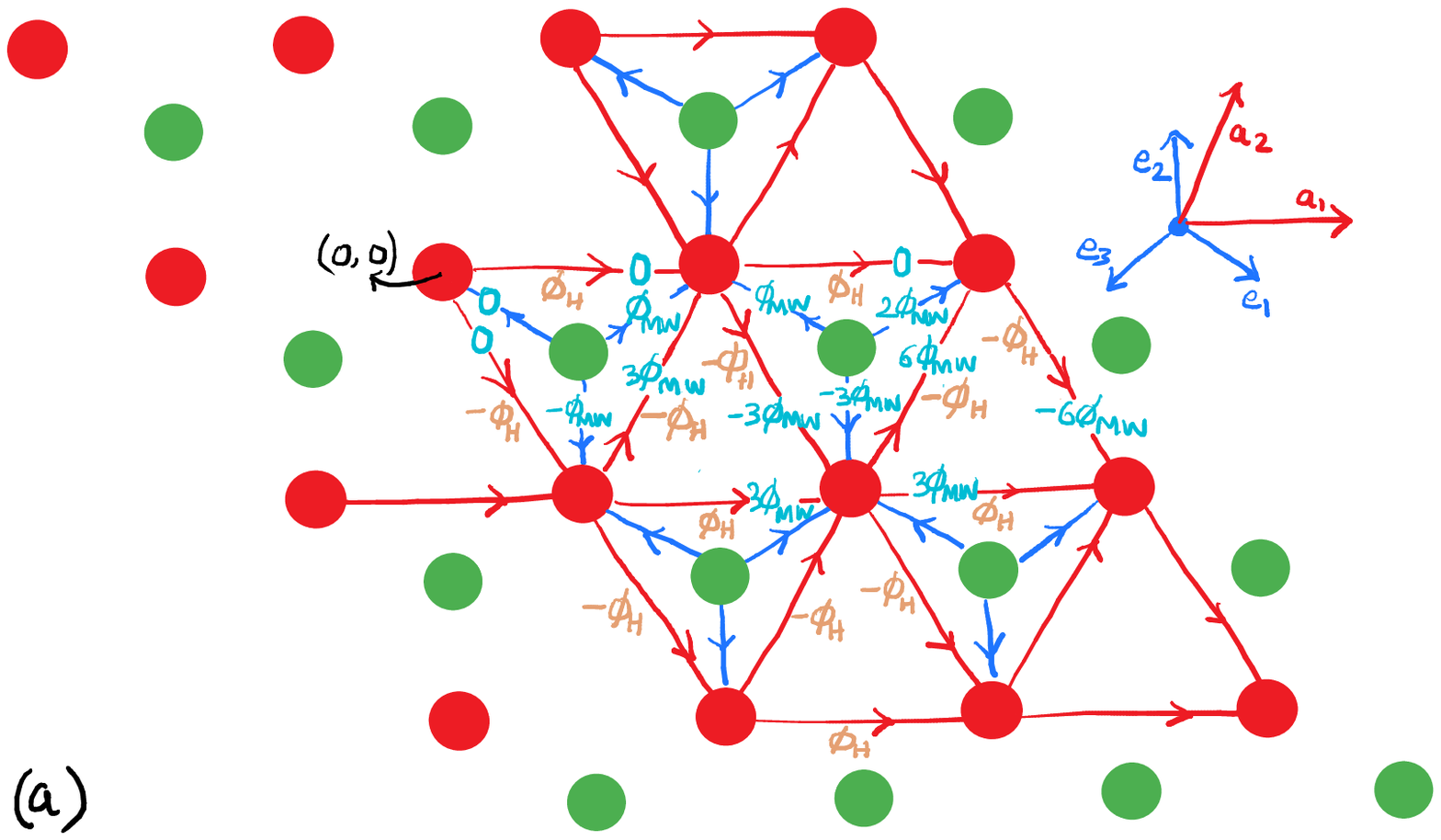}\\
\includegraphics[width=0.8\linewidth]{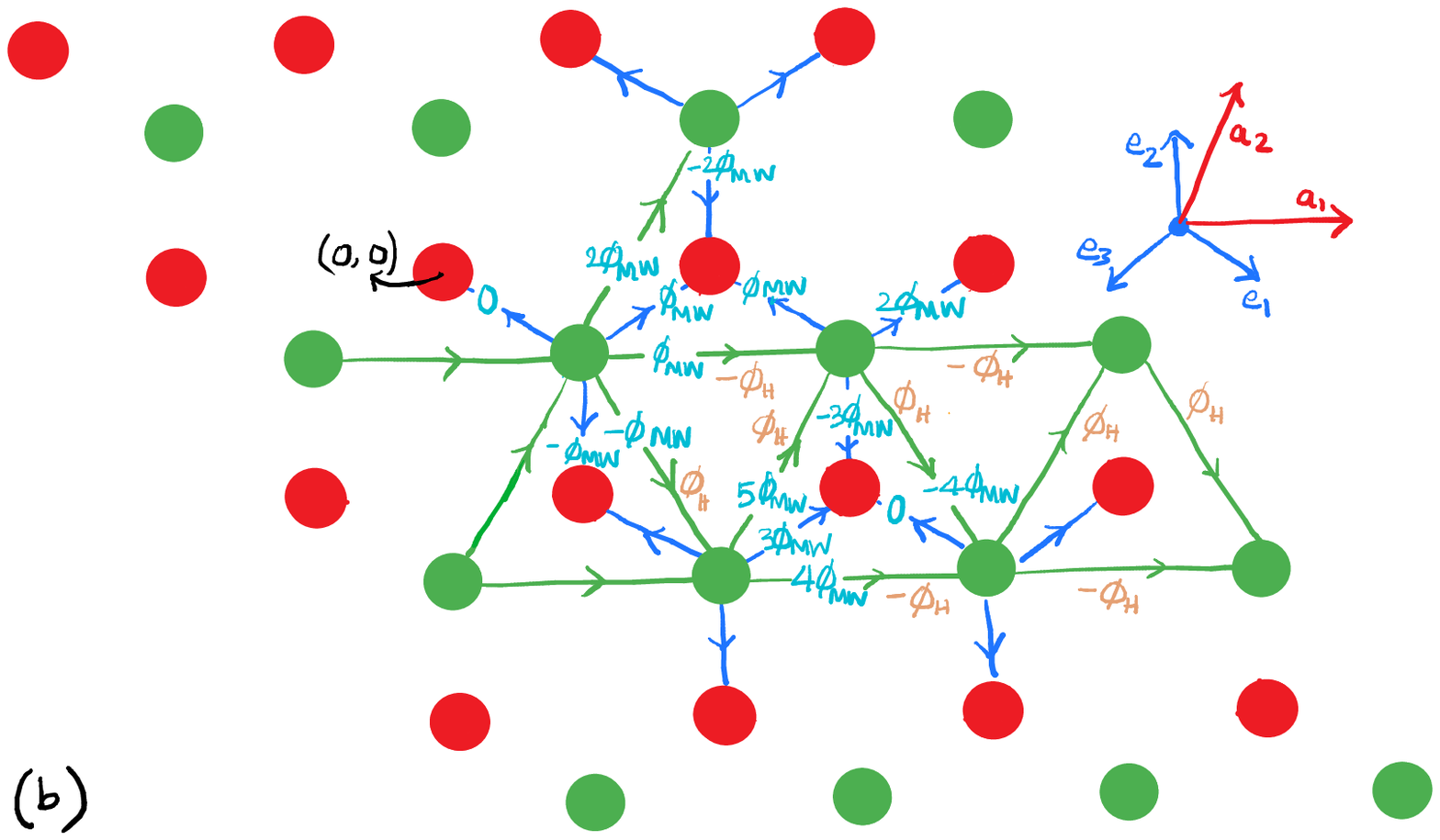}
\end{array}$\caption{
\label{fig:Xscheme} Extended scheme outlining the phase
accumulation in addition to $\int\vA\cdot d\mb{l}$ along each
bond. The field $\vA$ is written in symmetric gauge. The figure is
split into $(a)$ and $(b)$ for clarity. The flux $\phi_{MW}$ adds
up with area, whereas the flux $\phi_H$ denotes flux modulation
within the unit cell such that net contribution from $\phi_H$ to
the unit cell is zero. The presence of $\phi_H$ does not break
translational symmetry.}
\end{figure*}

For definiteness, we shall work in the symmetric gauge where
$\vA(\vrr)=(-y\hat\vx+x\hat{\mb{y}})B/2$. The phase accumulated on
a bond upon traversing in the direction $\mb{g}_n$ is given by
$\bar A_{g_n}\equiv\int_{\mb{g}_n}\vA\cdot d\mb{l}$. In the
lattice problem, we may treat $\vA$ to be constant for a unit
translation of $\mb{g}_n$, where
$\mb{g}_n\in\{\mb{e}_{1,2,3},\va_{1,2,3}\}$. The directions of
these vectors are illustrated in Fig. \ref{fig:Xscheme}. This
leads to $\bar A_{g_n}=A_{g_n}+(-yg_x+xg_y)Ba/2$, where $A_{g_n}$
refers to phase accumulation that can be gauged out and is the
same contribution from every lattice point. Thus, $A_{e_1}+A_{e_3}+A_{e_3}=0$ and
$A_{a_1}=A_{a_2}+A_{a_3}$. Once an origin is picked, we can start
assigning $\bar A^{a/b}$ to denote phase accumulation originating
from atom $a/b$ of the lattice. Thus for a given bond along $\mb{g}_n$, the phase can be written as $\bar A_{g_n}=\bar A^{a/b}_{g_n}+\Delta\bar A_{g_n}$, where $\bar A^{a/b}_{g_n}$ denotes the value of the bond at the origin, and $\Delta\bar A_{g_n}$ denotes the value at a bond separated from the origin by $(\Delta x,\Delta y)$. These quantities are expressed as (in the symmetric gauge):
\bea\label{eq:phaseAcc} \Delta\bar A_{a_1}&=&
-\phi_{MW}\left(\frac{2\sqrt{3}\Delta y}{a}\right),\\
\Delta\bar A_{a_2}&=&
\phi_{MW}\left(\frac{3\Delta x}{a}-\frac{\sqrt{3}\Delta y}{a}\right),\\
\Delta\bar A_{a_3}&=&
-\phi_{MW}\left(\frac{3\Delta x}{a}+\frac{\sqrt{3}\Delta y}{a}\right),\\
\Delta\bar A_{e_1}&=&-
\phi_{MW}\left(\frac{\Delta x}{a}+\frac{\sqrt{3}\Delta y}{a}\right),\\
\Delta\bar A_{e_2}&=&
\phi_{MW}\left(\frac{2\Delta x}{a}\right),\\
\Delta\bar A_{e_3}&=& -\phi_B\left(\frac{\Delta
x}{a}-\frac{\sqrt{3}\Delta y}{a}\right). \eea To initialize the bonds
starting from the $a/b$ atoms belonging to the lattice point at
the origin: \bea\label{eq:phaseAcc0}
\bar A^a_{a_{1}}&=&A_{a_{1}}+\phi_{H},\\
\bar A^a_{a_{2}}&=&A_{a_{2}}-\phi_{H},\\
\bar A^a_{a_{3}}&=&A_{a_{3}}-\phi_{H},\\
\bar A^b_{a_1}&=&A_{a_{1}}+\phi_{MW}-\phi_{H},\\
\bar A^b_{a_2}&=&A_{a_{2}}+2\phi_{MW}+\phi_{H},\\
\bar A^b_{a_3}&=&A_{a_{3}}-\phi_{MW}+\phi_{H},\\
\bar A^b_{e_1}&=&A_{e_{1}},\\
\bar A^b_{e_2}&=&A_{e_{2}}+\phi_{MW},\\
\bar A^b_{e_3}&=&A_{e_{3}}-\phi_{MW}.\eea And finally
$A_{g_n}\equiv\vA'\cdot \mb{g}_n$, where $\vA'$ is some vector
which will be gauged out and can be set to zero for our purposes.
The extended scheme is summarized in Fig. \ref{fig:Xscheme}. Note that any field that grows with area will be treated as $\phi_{MW}$ and the internal modulation will be treated as $\phi_H$.

\section{Effective Hamiltonians}
Following the prescription in the above section, we are able to assign the phases to every matrix element involved in the Hamiltonian. The Hamiltonian of nearest neighbor Graphene in an external field can be derived from
the lattice Hamiltonian as $H\rightarrow H_B$ where
\bea\label{eq:Helements} H^{11}_B&=&
r\left\{3+2\left[\cos(k_{0,1}-k_{0,2}+p_1-p_2)
\right.\right.\nonumber\\
&&~~+\cos(k_{0,2}-k_{0,3}+p_2-p_3)\nonumber\\
&&\left.\left. + \cos(k_{0,3}-k_{0,1}+p_3-p_1)\right]\right\},\nonumber\\
H^{22}_B&=& r\left\{3+2\left[\cos(k_{0,1}-k_{0,2}+p_1-p_2)
\right.\right.\nonumber\\
&&~~+\cos(k_{0,2}-k_{0,3}+p_2-p_3)\nonumber\\
&&\left.\left. + \cos(k_{0,3}-k_{0,1}+p_3-p_1)\right]\right\},\nonumber\\
H^{12}_B&=&T_{\vk_0+\vp},\nonumber\\
H^{21}_B&=&T^*_{\vk_0+\vp},\eea where $x_i\equiv\vx\cdot\mb{e}_i$, $\vk_0$ is a point about which the semi-classical momentum elongation is carried out. It is understood that $H_B$ is to be expanded to
$\mathcal{O}(p^2)$. Here the elongated momentum $vp$ involves $\vA^B$ such that $\mb{\nabla}\times\vA^B=\mb{B}$.

When we consider the case of a Chern-Simons'(CS) field, we have to account for $\phi_H$. The Hamiltonian matrix elements look like \bea\label{eq:Helements2} H^{11}_{CS}&=&
r\left\{3+2\left[\cos(k_{0,1}-k_{0,2}+p_1-p_2-e\phi_H)
\right.\right.\nonumber\\
&&~~+\cos(k_{0,2}-k_{0,3}+p_2-p_3-e\phi_H)\nonumber\\
&&\left.\left. + \cos(k_{0,3}-k_{0,1}+p_3-p_1-e\phi_{H})\right]\right\},\nonumber\\
H^{22}_{CS}&=&
r\left\{3+2\left[\cos(k_{0,1}-k_{0,2}+p_1-p_2+e\phi_H)
\right.\right.\nonumber\\
&&~~+\cos(k_{0,2}-k_{0,3}+p_2-p_3+e\phi_H)\nonumber\\
&&\left.\left. + \cos(k_{0,3}-k_{0,1}+p_3-p_1+e\phi_H)\right]\right\},\nonumber\\
H^{12}_{CS}&=&T_{\vk_0+\vp},\nonumber\\
H^{21}_{CS}&=&T^*_{\vk_0+\vp}.\eea Here the elongated momentum $vp$ involves $\vA^{MW}$ such that $\mb{\nabla}\times\vA^{MW}=\mb{B}_{MW}$. Note also that the $a/b$ atoms experience different signs of the flux $\phi_H$ although they originate from the same field $\vA^{MW}$.

To address the HLR case, we have to account for the external field and the CS field. This results in a very similar looking Hamiltonian as in Eq. (\ref{eq:Helements2}) but with $\vp$ containing $\vA^B$ and $\vA^{MW}$, and $\phi_H$ is still only generated from $\vA^{MW}$. The flux $\phi_B$ is generated from $\vA^B$ and the flux $\phi_{MW}$ is generated from $\vA^{MW}$.

\begin{figure}[htp]
\includegraphics[width=1\columnwidth]{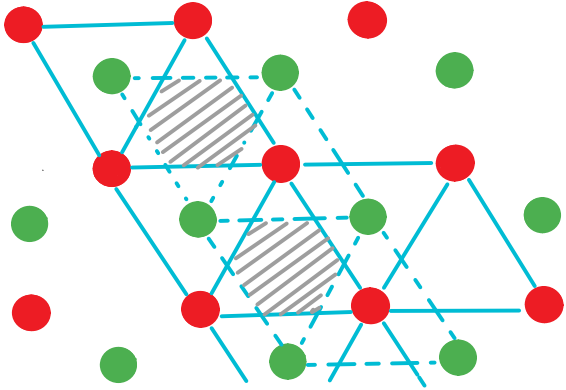}
\caption{ \label{fig:Xscheme2}The flux distribution arising from CS field in Graphene with next nearest neighbor hopping. The shaded region denotes the region of non-zero flux.}
\end{figure}

When $\phi_{b}=-\phi_{mw}=\phi_h$, we realize the Haldane's model of flux distribution in the unit cell. The extended scheme of the flux distribution shown in Fig. 1 of the main text (MT), is shown in Fig. \ref{fig:Xscheme2} in this text.

%\end{appendices}

\bibliography{Ref}

\begin{thebibliography}{45}
\expandafter\ifx\csname natexlab\endcsname\relax\def\natexlab#1{#1}\fi
\expandafter\ifx\csname bibnamefont\endcsname\relax
  \def\bibnamefont#1{#1}\fi
\expandafter\ifx\csname bibfnamefont\endcsname\relax
  \def\bibfnamefont#1{#1}\fi
\expandafter\ifx\csname citenamefont\endcsname\relax
  \def\citenamefont#1{#1}\fi
\expandafter\ifx\csname url\endcsname\relax
  \def\url#1{\texttt{#1}}\fi
\expandafter\ifx\csname urlprefix\endcsname\relax\def\urlprefix{URL }\fi
\providecommand{\bibinfo}[2]{#2}
\providecommand{\eprint}[2][]{\url{#2}}

\bibitem[{\citenamefont{Tsui et~al.}(1982)\citenamefont{Tsui, Stormer, and
  Gossard}}]{Tsui1983}
\bibinfo{author}{\bibfnamefont{D.~C.} \bibnamefont{Tsui}},
  \bibinfo{author}{\bibfnamefont{H.~L.} \bibnamefont{Stormer}},
  \bibnamefont{and} \bibinfo{author}{\bibfnamefont{A.~C.}
  \bibnamefont{Gossard}}, \bibinfo{journal}{Phys. Rev. Lett.}
  \textbf{\bibinfo{volume}{48}}, \bibinfo{pages}{1559} (\bibinfo{year}{1982}),
  \urlprefix\url{https://link.aps.org/doi/10.1103/PhysRevLett.48.1559}.

\bibitem[{\citenamefont{Kukushkin et~al.}(1999)\citenamefont{Kukushkin,
  v.~Klitzing, and Eberl}}]{Kukushkin1999}
\bibinfo{author}{\bibfnamefont{I.~V.} \bibnamefont{Kukushkin}},
  \bibinfo{author}{\bibfnamefont{K.}~\bibnamefont{v.~Klitzing}},
  \bibnamefont{and} \bibinfo{author}{\bibfnamefont{K.}~\bibnamefont{Eberl}},
  \bibinfo{journal}{Phys. Rev. Lett.} \textbf{\bibinfo{volume}{82}},
  \bibinfo{pages}{3665} (\bibinfo{year}{1999}),
  \urlprefix\url{https://link.aps.org/doi/10.1103/PhysRevLett.82.3665}.

\bibitem[{\citenamefont{Verdene et~al.}(2007)\citenamefont{Verdene, Martin,
  Gamez, Smet, von Klitzing, Mahalu, Schuh, Abstreiter, and
  Yacoby}}]{Verdene2007}
\bibinfo{author}{\bibfnamefont{B.}~\bibnamefont{Verdene}},
  \bibinfo{author}{\bibfnamefont{J.}~\bibnamefont{Martin}},
  \bibinfo{author}{\bibfnamefont{G.}~\bibnamefont{Gamez}},
  \bibinfo{author}{\bibfnamefont{J.}~\bibnamefont{Smet}},
  \bibinfo{author}{\bibfnamefont{K.}~\bibnamefont{von Klitzing}},
  \bibinfo{author}{\bibfnamefont{D.}~\bibnamefont{Mahalu}},
  \bibinfo{author}{\bibfnamefont{D.}~\bibnamefont{Schuh}},
  \bibinfo{author}{\bibfnamefont{G.}~\bibnamefont{Abstreiter}},
  \bibnamefont{and} \bibinfo{author}{\bibfnamefont{A.}~\bibnamefont{Yacoby}},
  \bibinfo{journal}{Nature Physics} \textbf{\bibinfo{volume}{3}},
  \bibinfo{pages}{392 EP } (\bibinfo{year}{2007}),
  \urlprefix\url{https://doi.org/10.1038/nphys588}.

\bibitem[{\citenamefont{Dean et~al.}(2011)\citenamefont{Dean, Young,
  Cadden-Zimansky, Wang, Ren, Watanabe, Taniguchi, Kim, Hone, and
  Shepard}}]{Dean2011}
\bibinfo{author}{\bibfnamefont{C.~R.} \bibnamefont{Dean}},
  \bibinfo{author}{\bibfnamefont{A.~F.} \bibnamefont{Young}},
  \bibinfo{author}{\bibfnamefont{P.}~\bibnamefont{Cadden-Zimansky}},
  \bibinfo{author}{\bibfnamefont{L.}~\bibnamefont{Wang}},
  \bibinfo{author}{\bibfnamefont{H.}~\bibnamefont{Ren}},
  \bibinfo{author}{\bibfnamefont{K.}~\bibnamefont{Watanabe}},
  \bibinfo{author}{\bibfnamefont{T.}~\bibnamefont{Taniguchi}},
  \bibinfo{author}{\bibfnamefont{P.}~\bibnamefont{Kim}},
  \bibinfo{author}{\bibfnamefont{J.}~\bibnamefont{Hone}}, \bibnamefont{and}
  \bibinfo{author}{\bibfnamefont{K.~L.} \bibnamefont{Shepard}},
  \bibinfo{journal}{Nature Physics} \textbf{\bibinfo{volume}{7}},
  \bibinfo{pages}{693 EP } (\bibinfo{year}{2011}),
  \urlprefix\url{https://doi.org/10.1038/nphys2007}.

\bibitem[{\citenamefont{Du et~al.}(2009)\citenamefont{Du, Skachko, Duerr,
  Luican, and Andrei}}]{Du2009}
\bibinfo{author}{\bibfnamefont{X.}~\bibnamefont{Du}},
  \bibinfo{author}{\bibfnamefont{I.}~\bibnamefont{Skachko}},
  \bibinfo{author}{\bibfnamefont{F.}~\bibnamefont{Duerr}},
  \bibinfo{author}{\bibfnamefont{A.}~\bibnamefont{Luican}}, \bibnamefont{and}
  \bibinfo{author}{\bibfnamefont{E.~Y.} \bibnamefont{Andrei}},
  \bibinfo{journal}{Nature} \textbf{\bibinfo{volume}{462}}, \bibinfo{pages}{192
  EP } (\bibinfo{year}{2009}),
  \urlprefix\url{https://doi.org/10.1038/nature08522}.

\bibitem[{\citenamefont{Bolotin et~al.}(2009)\citenamefont{Bolotin, Ghahari,
  Shulman, Stormer, and Kim}}]{Bolotin2009}
\bibinfo{author}{\bibfnamefont{K.~I.} \bibnamefont{Bolotin}},
  \bibinfo{author}{\bibfnamefont{F.}~\bibnamefont{Ghahari}},
  \bibinfo{author}{\bibfnamefont{M.~D.} \bibnamefont{Shulman}},
  \bibinfo{author}{\bibfnamefont{H.~L.} \bibnamefont{Stormer}},
  \bibnamefont{and} \bibinfo{author}{\bibfnamefont{P.}~\bibnamefont{Kim}},
  \bibinfo{journal}{Nature} \textbf{\bibinfo{volume}{462}}, \bibinfo{pages}{196
  EP } (\bibinfo{year}{2009}),
  \urlprefix\url{https://doi.org/10.1038/nature08582}.

\bibitem[{\citenamefont{Ghahari et~al.}(2011)\citenamefont{Ghahari, Zhao,
  Cadden-Zimansky, Bolotin, and Kim}}]{Ghahari2011}
\bibinfo{author}{\bibfnamefont{F.}~\bibnamefont{Ghahari}},
  \bibinfo{author}{\bibfnamefont{Y.}~\bibnamefont{Zhao}},
  \bibinfo{author}{\bibfnamefont{P.}~\bibnamefont{Cadden-Zimansky}},
  \bibinfo{author}{\bibfnamefont{K.}~\bibnamefont{Bolotin}}, \bibnamefont{and}
  \bibinfo{author}{\bibfnamefont{P.}~\bibnamefont{Kim}},
  \bibinfo{journal}{Phys. Rev. Lett.} \textbf{\bibinfo{volume}{106}},
  \bibinfo{pages}{046801} (\bibinfo{year}{2011}),
  \urlprefix\url{https://link.aps.org/doi/10.1103/PhysRevLett.106.046801}.

\bibitem[{\citenamefont{Feldman et~al.}(2012)\citenamefont{Feldman, Krauss,
  Smet, and Yacoby}}]{Feldman2012}
\bibinfo{author}{\bibfnamefont{B.~E.} \bibnamefont{Feldman}},
  \bibinfo{author}{\bibfnamefont{B.}~\bibnamefont{Krauss}},
  \bibinfo{author}{\bibfnamefont{J.~H.} \bibnamefont{Smet}}, \bibnamefont{and}
  \bibinfo{author}{\bibfnamefont{A.}~\bibnamefont{Yacoby}},
  \bibinfo{journal}{Science} \textbf{\bibinfo{volume}{337}},
  \bibinfo{pages}{1196} (\bibinfo{year}{2012}), ISSN \bibinfo{issn}{0036-8075},
  \eprint{http://science.sciencemag.org/content/337/6099/1196.full.pdf},
  \urlprefix\url{http://science.sciencemag.org/content/337/6099/1196}.

\bibitem[{\citenamefont{Feldman et~al.}(2013)\citenamefont{Feldman, Levin,
  Krauss, Abanin, Halperin, Smet, and Yacoby}}]{Feldman2013}
\bibinfo{author}{\bibfnamefont{B.~E.} \bibnamefont{Feldman}},
  \bibinfo{author}{\bibfnamefont{A.~J.} \bibnamefont{Levin}},
  \bibinfo{author}{\bibfnamefont{B.}~\bibnamefont{Krauss}},
  \bibinfo{author}{\bibfnamefont{D.~A.} \bibnamefont{Abanin}},
  \bibinfo{author}{\bibfnamefont{B.~I.} \bibnamefont{Halperin}},
  \bibinfo{author}{\bibfnamefont{J.~H.} \bibnamefont{Smet}}, \bibnamefont{and}
  \bibinfo{author}{\bibfnamefont{A.}~\bibnamefont{Yacoby}},
  \bibinfo{journal}{Phys. Rev. Lett.} \textbf{\bibinfo{volume}{111}},
  \bibinfo{pages}{076802} (\bibinfo{year}{2013}),
  \urlprefix\url{https://link.aps.org/doi/10.1103/PhysRevLett.111.076802}.

\bibitem[{\citenamefont{Jain}(1989)}]{Jain1989}
\bibinfo{author}{\bibfnamefont{J.~K.} \bibnamefont{Jain}},
  \bibinfo{journal}{Phys. Rev. Lett.} \textbf{\bibinfo{volume}{63}},
  \bibinfo{pages}{199} (\bibinfo{year}{1989}),
  \urlprefix\url{https://link.aps.org/doi/10.1103/PhysRevLett.63.199}.

\bibitem[{\citenamefont{Zibrov et~al.}(2017)\citenamefont{Zibrov, Kometter,
  Zhou, Spanton, Taniguchi, Watanabe, Zaletel, and Young}}]{Zibrov2017}
\bibinfo{author}{\bibfnamefont{A.~A.} \bibnamefont{Zibrov}},
  \bibinfo{author}{\bibfnamefont{C.}~\bibnamefont{Kometter}},
  \bibinfo{author}{\bibfnamefont{H.}~\bibnamefont{Zhou}},
  \bibinfo{author}{\bibfnamefont{E.~M.} \bibnamefont{Spanton}},
  \bibinfo{author}{\bibfnamefont{T.}~\bibnamefont{Taniguchi}},
  \bibinfo{author}{\bibfnamefont{K.}~\bibnamefont{Watanabe}},
  \bibinfo{author}{\bibfnamefont{M.~P.} \bibnamefont{Zaletel}},
  \bibnamefont{and} \bibinfo{author}{\bibfnamefont{A.~F.} \bibnamefont{Young}},
  \bibinfo{journal}{Nature} \textbf{\bibinfo{volume}{549}}, \bibinfo{pages}{360
  EP } (\bibinfo{year}{2017}),
  \urlprefix\url{https://doi.org/10.1038/nature23893}.

\bibitem[{\citenamefont{Laughlin}(1983)}]{Laughlin1983}
\bibinfo{author}{\bibfnamefont{R.~B.} \bibnamefont{Laughlin}},
  \bibinfo{journal}{Phys. Rev. Lett.} \textbf{\bibinfo{volume}{50}},
  \bibinfo{pages}{1395} (\bibinfo{year}{1983}),
  \urlprefix\url{https://link.aps.org/doi/10.1103/PhysRevLett.50.1395}.

\bibitem[{\citenamefont{Jain}(2015)}]{Jain2015}
\bibinfo{author}{\bibfnamefont{J.~K.} \bibnamefont{Jain}},
  \bibinfo{journal}{Annual Review of Condensed Matter Physics}
  \textbf{\bibinfo{volume}{6}}, \bibinfo{pages}{39} (\bibinfo{year}{2015}),
  \eprint{https://doi.org/10.1146/annurev-conmatphys-031214-014606},
  \urlprefix\url{https://doi.org/10.1146/annurev-conmatphys-031214-014606}.

\bibitem[{\citenamefont{Zhang et~al.}(1989)\citenamefont{Zhang, Hansson, and
  Kivelson}}]{Zhang1989}
\bibinfo{author}{\bibfnamefont{S.~C.} \bibnamefont{Zhang}},
  \bibinfo{author}{\bibfnamefont{T.~H.} \bibnamefont{Hansson}},
  \bibnamefont{and} \bibinfo{author}{\bibfnamefont{S.}~\bibnamefont{Kivelson}},
  \bibinfo{journal}{Phys. Rev. Lett.} \textbf{\bibinfo{volume}{62}},
  \bibinfo{pages}{82} (\bibinfo{year}{1989}),
  \urlprefix\url{https://link.aps.org/doi/10.1103/PhysRevLett.62.82}.

\bibitem[{\citenamefont{Lopez and Fradkin}(1991)}]{Lopez1991}
\bibinfo{author}{\bibfnamefont{A.}~\bibnamefont{Lopez}} \bibnamefont{and}
  \bibinfo{author}{\bibfnamefont{E.}~\bibnamefont{Fradkin}},
  \bibinfo{journal}{Phys. Rev. B} \textbf{\bibinfo{volume}{44}},
  \bibinfo{pages}{5246} (\bibinfo{year}{1991}),
  \urlprefix\url{https://link.aps.org/doi/10.1103/PhysRevB.44.5246}.

\bibitem[{\citenamefont{Lopez and Fradkin}(1995)}]{Lopez1995}
\bibinfo{author}{\bibfnamefont{A.}~\bibnamefont{Lopez}} \bibnamefont{and}
  \bibinfo{author}{\bibfnamefont{E.}~\bibnamefont{Fradkin}},
  \bibinfo{journal}{Phys. Rev. B} \textbf{\bibinfo{volume}{51}},
  \bibinfo{pages}{4347} (\bibinfo{year}{1995}),
  \urlprefix\url{https://link.aps.org/doi/10.1103/PhysRevB.51.4347}.

\bibitem[{\citenamefont{Wen}(1995)}]{Wen1995}
\bibinfo{author}{\bibfnamefont{X.-G.} \bibnamefont{Wen}},
  \bibinfo{journal}{Advances in Physics} \textbf{\bibinfo{volume}{44}},
  \bibinfo{pages}{405} (\bibinfo{year}{1995}), ISSN \bibinfo{issn}{0001-8732},
  \urlprefix\url{https://doi.org/10.1080/00018739500101566}.

\bibitem[{\citenamefont{Davenport and Simon}(2012)}]{Davenport2012}
\bibinfo{author}{\bibfnamefont{S.~C.} \bibnamefont{Davenport}}
  \bibnamefont{and} \bibinfo{author}{\bibfnamefont{S.~H.} \bibnamefont{Simon}},
  \bibinfo{journal}{Phys. Rev. B} \textbf{\bibinfo{volume}{85}},
  \bibinfo{pages}{245303} (\bibinfo{year}{2012}),
  \urlprefix\url{https://link.aps.org/doi/10.1103/PhysRevB.85.245303}.

\bibitem[{\citenamefont{Halperin et~al.}(1993)\citenamefont{Halperin, Lee, and
  Read}}]{Halperin1993}
\bibinfo{author}{\bibfnamefont{B.~I.} \bibnamefont{Halperin}},
  \bibinfo{author}{\bibfnamefont{P.~A.} \bibnamefont{Lee}}, \bibnamefont{and}
  \bibinfo{author}{\bibfnamefont{N.}~\bibnamefont{Read}},
  \bibinfo{journal}{Phys. Rev. B} \textbf{\bibinfo{volume}{47}},
  \bibinfo{pages}{7312} (\bibinfo{year}{1993}),
  \urlprefix\url{https://link.aps.org/doi/10.1103/PhysRevB.47.7312}.

\bibitem[{\citenamefont{Son}(2015)}]{Son2015}
\bibinfo{author}{\bibfnamefont{D.~T.} \bibnamefont{Son}},
  \bibinfo{journal}{Phys. Rev. X} \textbf{\bibinfo{volume}{5}},
  \bibinfo{pages}{031027} (\bibinfo{year}{2015}),
  \urlprefix\url{https://link.aps.org/doi/10.1103/PhysRevX.5.031027}.

\bibitem[{\citenamefont{Goldman and Fradkin}(2018)}]{Goldman2018}
\bibinfo{author}{\bibfnamefont{H.}~\bibnamefont{Goldman}} \bibnamefont{and}
  \bibinfo{author}{\bibfnamefont{E.}~\bibnamefont{Fradkin}},
  \bibinfo{journal}{Phys. Rev. B} \textbf{\bibinfo{volume}{98}},
  \bibinfo{pages}{165137} (\bibinfo{year}{2018}),
  \urlprefix\url{https://link.aps.org/doi/10.1103/PhysRevB.98.165137}.

\bibitem[{\citenamefont{You}(2018)}]{You2018}
\bibinfo{author}{\bibfnamefont{Y.}~\bibnamefont{You}}, \bibinfo{journal}{Phys.
  Rev. B} \textbf{\bibinfo{volume}{97}}, \bibinfo{pages}{165115}
  (\bibinfo{year}{2018}),
  \urlprefix\url{https://link.aps.org/doi/10.1103/PhysRevB.97.165115}.

\bibitem[{\citenamefont{Sedrakyan and Raikh}(2008)}]{Sedrakyan2008}
\bibinfo{author}{\bibfnamefont{T.~A.} \bibnamefont{Sedrakyan}}
  \bibnamefont{and} \bibinfo{author}{\bibfnamefont{M.~E.} \bibnamefont{Raikh}},
  \bibinfo{journal}{Phys. Rev. B} \textbf{\bibinfo{volume}{77}},
  \bibinfo{pages}{115353} (\bibinfo{year}{2008}),
  \urlprefix\url{https://link.aps.org/doi/10.1103/PhysRevB.77.115353}.

\bibitem[{\citenamefont{Sedrakyan and Chubukov}(2009)}]{Sedrakyan2009}
\bibinfo{author}{\bibfnamefont{T.~A.} \bibnamefont{Sedrakyan}}
  \bibnamefont{and} \bibinfo{author}{\bibfnamefont{A.~V.}
  \bibnamefont{Chubukov}}, \bibinfo{journal}{Phys. Rev. B}
  \textbf{\bibinfo{volume}{79}}, \bibinfo{pages}{115129}
  (\bibinfo{year}{2009}),
  \urlprefix\url{https://link.aps.org/doi/10.1103/PhysRevB.79.115129}.

\bibitem[{\citenamefont{Geraedts et~al.}(2016)\citenamefont{Geraedts, Zaletel,
  Mong, Metlitski, Vishwanath, and Motrunich}}]{Geraedts2016}
\bibinfo{author}{\bibfnamefont{S.~D.} \bibnamefont{Geraedts}},
  \bibinfo{author}{\bibfnamefont{M.~P.} \bibnamefont{Zaletel}},
  \bibinfo{author}{\bibfnamefont{R.~S.~K.} \bibnamefont{Mong}},
  \bibinfo{author}{\bibfnamefont{M.~A.} \bibnamefont{Metlitski}},
  \bibinfo{author}{\bibfnamefont{A.}~\bibnamefont{Vishwanath}},
  \bibnamefont{and} \bibinfo{author}{\bibfnamefont{O.~I.}
  \bibnamefont{Motrunich}}, \bibinfo{journal}{Science}
  \textbf{\bibinfo{volume}{352}}, \bibinfo{pages}{197} (\bibinfo{year}{2016}),
  ISSN \bibinfo{issn}{0036-8075},
  \eprint{http://science.sciencemag.org/content/352/6282/197.full.pdf},
  \urlprefix\url{http://science.sciencemag.org/content/352/6282/197}.

\bibitem[{\citenamefont{Wang et~al.}(2017)\citenamefont{Wang, Cooper, Halperin,
  and Stern}}]{Wang2017X}
\bibinfo{author}{\bibfnamefont{C.}~\bibnamefont{Wang}},
  \bibinfo{author}{\bibfnamefont{N.~R.} \bibnamefont{Cooper}},
  \bibinfo{author}{\bibfnamefont{B.~I.} \bibnamefont{Halperin}},
  \bibnamefont{and} \bibinfo{author}{\bibfnamefont{A.}~\bibnamefont{Stern}},
  \bibinfo{journal}{Phys. Rev. X} \textbf{\bibinfo{volume}{7}},
  \bibinfo{pages}{031029} (\bibinfo{year}{2017}),
  \urlprefix\url{https://link.aps.org/doi/10.1103/PhysRevX.7.031029}.

\bibitem[{\citenamefont{Kumar et~al.}(2014)\citenamefont{Kumar, Sun, and
  Fradkin}}]{Kumar2014}
\bibinfo{author}{\bibfnamefont{K.}~\bibnamefont{Kumar}},
  \bibinfo{author}{\bibfnamefont{K.}~\bibnamefont{Sun}}, \bibnamefont{and}
  \bibinfo{author}{\bibfnamefont{E.}~\bibnamefont{Fradkin}},
  \bibinfo{journal}{Phys. Rev. B} \textbf{\bibinfo{volume}{90}},
  \bibinfo{pages}{174409} (\bibinfo{year}{2014}),
  \urlprefix\url{https://link.aps.org/doi/10.1103/PhysRevB.90.174409}.

\bibitem[{\citenamefont{Sedrakyan et~al.}(2012)\citenamefont{Sedrakyan,
  Kamenev, and Glazman}}]{Sedrakyan2012}
\bibinfo{author}{\bibfnamefont{T.~A.} \bibnamefont{Sedrakyan}},
  \bibinfo{author}{\bibfnamefont{A.}~\bibnamefont{Kamenev}}, \bibnamefont{and}
  \bibinfo{author}{\bibfnamefont{L.~I.} \bibnamefont{Glazman}},
  \bibinfo{journal}{Phys. Rev. A} \textbf{\bibinfo{volume}{86}},
  \bibinfo{pages}{063639} (\bibinfo{year}{2012}),
  \urlprefix\url{https://link.aps.org/doi/10.1103/PhysRevA.86.063639}.

\bibitem[{\citenamefont{Sedrakyan et~al.}(2014)\citenamefont{Sedrakyan,
  Glazman, and Kamenev}}]{Sedrakyan2014}
\bibinfo{author}{\bibfnamefont{T.~A.} \bibnamefont{Sedrakyan}},
  \bibinfo{author}{\bibfnamefont{L.~I.} \bibnamefont{Glazman}},
  \bibnamefont{and} \bibinfo{author}{\bibfnamefont{A.}~\bibnamefont{Kamenev}},
  \bibinfo{journal}{Phys. Rev. B} \textbf{\bibinfo{volume}{89}},
  \bibinfo{pages}{201112} (\bibinfo{year}{2014}),
  \urlprefix\url{https://link.aps.org/doi/10.1103/PhysRevB.89.201112}.

\bibitem[{\citenamefont{Sedrakyan
  et~al.}(2015{\natexlab{a}})\citenamefont{Sedrakyan, Glazman, and
  Kamenev}}]{Sedrakyan2015}
\bibinfo{author}{\bibfnamefont{T.~A.} \bibnamefont{Sedrakyan}},
  \bibinfo{author}{\bibfnamefont{L.~I.} \bibnamefont{Glazman}},
  \bibnamefont{and} \bibinfo{author}{\bibfnamefont{A.}~\bibnamefont{Kamenev}},
  \bibinfo{journal}{Phys. Rev. Lett.} \textbf{\bibinfo{volume}{114}},
  \bibinfo{pages}{037203} (\bibinfo{year}{2015}{\natexlab{a}}),
  \urlprefix\url{https://link.aps.org/doi/10.1103/PhysRevLett.114.037203}.

\bibitem[{\citenamefont{Maiti and Sedrakyan}(2019)}]{Maiti2018}
\bibinfo{author}{\bibfnamefont{S.}~\bibnamefont{Maiti}} \bibnamefont{and}
  \bibinfo{author}{\bibfnamefont{T.}~\bibnamefont{Sedrakyan}},
  \bibinfo{journal}{Phys. Rev. B} \textbf{\bibinfo{volume}{99}},
  \bibinfo{pages}{174418} (\bibinfo{year}{2019}),
  \urlprefix\url{https://link.aps.org/doi/10.1103/PhysRevB.99.174418}.

\bibitem[{\citenamefont{Sodemann and MacDonald}(2014)}]{Sodeman2014}
\bibinfo{author}{\bibfnamefont{I.}~\bibnamefont{Sodemann}} \bibnamefont{and}
  \bibinfo{author}{\bibfnamefont{A.~H.} \bibnamefont{MacDonald}},
  \bibinfo{journal}{Phys. Rev. Lett.} \textbf{\bibinfo{volume}{112}},
  \bibinfo{pages}{126804} (\bibinfo{year}{2014}),
  \urlprefix\url{https://link.aps.org/doi/10.1103/PhysRevLett.112.126804}.

\bibitem[{\citenamefont{Wu et~al.}(2015)\citenamefont{Wu, Sodemann, MacDonald,
  and Jolicoeur}}]{Wu2015}
\bibinfo{author}{\bibfnamefont{F.}~\bibnamefont{Wu}},
  \bibinfo{author}{\bibfnamefont{I.}~\bibnamefont{Sodemann}},
  \bibinfo{author}{\bibfnamefont{A.~H.} \bibnamefont{MacDonald}},
  \bibnamefont{and}
  \bibinfo{author}{\bibfnamefont{T.}~\bibnamefont{Jolicoeur}},
  \bibinfo{journal}{Phys. Rev. Lett.} \textbf{\bibinfo{volume}{115}},
  \bibinfo{pages}{166805} (\bibinfo{year}{2015}),
  \urlprefix\url{https://link.aps.org/doi/10.1103/PhysRevLett.115.166805}.

\bibitem[{\citenamefont{Haldane}(1988)}]{Haldane1988}
\bibinfo{author}{\bibfnamefont{F.~D.~M.} \bibnamefont{Haldane}},
  \bibinfo{journal}{Phys. Rev. Lett.} \textbf{\bibinfo{volume}{61}},
  \bibinfo{pages}{2015} (\bibinfo{year}{1988}),
  \urlprefix\url{https://link.aps.org/doi/10.1103/PhysRevLett.61.2015}.

\bibitem[{\citenamefont{Castro~Neto et~al.}(2009)\citenamefont{Castro~Neto,
  Guinea, Peres, Novoselov, and Geim}}]{CastroNeto2009}
\bibinfo{author}{\bibfnamefont{A.~H.} \bibnamefont{Castro~Neto}},
  \bibinfo{author}{\bibfnamefont{F.}~\bibnamefont{Guinea}},
  \bibinfo{author}{\bibfnamefont{N.~M.~R.} \bibnamefont{Peres}},
  \bibinfo{author}{\bibfnamefont{K.~S.} \bibnamefont{Novoselov}},
  \bibnamefont{and} \bibinfo{author}{\bibfnamefont{A.~K.} \bibnamefont{Geim}},
  \bibinfo{journal}{Rev. Mod. Phys.} \textbf{\bibinfo{volume}{81}},
  \bibinfo{pages}{109} (\bibinfo{year}{2009}),
  \urlprefix\url{https://link.aps.org/doi/10.1103/RevModPhys.81.109}.

\bibitem[{\citenamefont{Kundu}(2011)}]{Kundu2011}
\bibinfo{author}{\bibfnamefont{R.}~\bibnamefont{Kundu}}, \bibinfo{journal}{Mod.
  Phys. Lett. B} \textbf{\bibinfo{volume}{25}}, \bibinfo{pages}{163}
  (\bibinfo{year}{2011}),
  \urlprefix\url{https://www.worldscientific.com/doi/abs/10.1142/S0217984911025663}.

\bibitem[{\citenamefont{Hofstadter}(1976)}]{Hofstadter1976}
\bibinfo{author}{\bibfnamefont{D.~R.} \bibnamefont{Hofstadter}},
  \bibinfo{journal}{Phys. Rev. B} \textbf{\bibinfo{volume}{14}},
  \bibinfo{pages}{2239} (\bibinfo{year}{1976}),
  \urlprefix\url{https://link.aps.org/doi/10.1103/PhysRevB.14.2239}.

\bibitem[{\citenamefont{Sedrakyan
  et~al.}(2015{\natexlab{b}})\citenamefont{Sedrakyan, Galitski, and
  Kamenev}}]{Sedrakyan2015_2}
\bibinfo{author}{\bibfnamefont{T.~A.} \bibnamefont{Sedrakyan}},
  \bibinfo{author}{\bibfnamefont{V.~M.} \bibnamefont{Galitski}},
  \bibnamefont{and} \bibinfo{author}{\bibfnamefont{A.}~\bibnamefont{Kamenev}},
  \bibinfo{journal}{Phys. Rev. Lett.} \textbf{\bibinfo{volume}{115}},
  \bibinfo{pages}{195301} (\bibinfo{year}{2015}{\natexlab{b}}),
  \urlprefix\url{https://link.aps.org/doi/10.1103/PhysRevLett.115.195301}.

\bibitem[{\citenamefont{Narevich et~al.}(2001)\citenamefont{Narevich, Murthy,
  and Fertig}}]{Naverich2001}
\bibinfo{author}{\bibfnamefont{R.}~\bibnamefont{Narevich}},
  \bibinfo{author}{\bibfnamefont{G.}~\bibnamefont{Murthy}}, \bibnamefont{and}
  \bibinfo{author}{\bibfnamefont{H.~A.} \bibnamefont{Fertig}},
  \bibinfo{journal}{Phys. Rev. B} \textbf{\bibinfo{volume}{64}},
  \bibinfo{pages}{245326} (\bibinfo{year}{2001}),
  \urlprefix\url{https://link.aps.org/doi/10.1103/PhysRevB.64.245326}.

\bibitem[{\citenamefont{Shytov et~al.}(1998)\citenamefont{Shytov, Levitov, and
  Halperin}}]{PhysRevLett.80.141}
\bibinfo{author}{\bibfnamefont{A.~V.} \bibnamefont{Shytov}},
  \bibinfo{author}{\bibfnamefont{L.~S.} \bibnamefont{Levitov}},
  \bibnamefont{and} \bibinfo{author}{\bibfnamefont{B.~I.}
  \bibnamefont{Halperin}}, \bibinfo{journal}{Phys. Rev. Lett.}
  \textbf{\bibinfo{volume}{80}}, \bibinfo{pages}{141} (\bibinfo{year}{1998}),
  \urlprefix\url{https://link.aps.org/doi/10.1103/PhysRevLett.80.141}.

\bibitem[{\citenamefont{Str\"om and
  Johannesson}(2009)}]{PhysRevLett.102.096806}
\bibinfo{author}{\bibfnamefont{A.}~\bibnamefont{Str\"om}} \bibnamefont{and}
  \bibinfo{author}{\bibfnamefont{H.}~\bibnamefont{Johannesson}},
  \bibinfo{journal}{Phys. Rev. Lett.} \textbf{\bibinfo{volume}{102}},
  \bibinfo{pages}{096806} (\bibinfo{year}{2009}),
  \urlprefix\url{https://link.aps.org/doi/10.1103/PhysRevLett.102.096806}.

\bibitem[{\citenamefont{Kamenev and Gefen}(2015)}]{PhysRevLett.114.156401}
\bibinfo{author}{\bibfnamefont{A.}~\bibnamefont{Kamenev}} \bibnamefont{and}
  \bibinfo{author}{\bibfnamefont{Y.}~\bibnamefont{Gefen}},
  \bibinfo{journal}{Phys. Rev. Lett.} \textbf{\bibinfo{volume}{114}},
  \bibinfo{pages}{156401} (\bibinfo{year}{2015}),
  \urlprefix\url{https://link.aps.org/doi/10.1103/PhysRevLett.114.156401}.

\bibitem[{\citenamefont{Kane and Fisher}(1997)}]{fisher}
\bibinfo{author}{\bibfnamefont{C.~L.} \bibnamefont{Kane}} \bibnamefont{and}
  \bibinfo{author}{\bibfnamefont{M.~P.~A.} \bibnamefont{Fisher}},
  \bibinfo{journal}{Phys. Rev. B} \textbf{\bibinfo{volume}{55}},
  \bibinfo{pages}{15832} (\bibinfo{year}{1997}),
  \urlprefix\url{https://link.aps.org/doi/10.1103/PhysRevB.55.15832}.

\bibitem[{\citenamefont{Fern et~al.}(2018)\citenamefont{Fern, Bondesan, and
  Simon}}]{PhysRevB.98.155321}
\bibinfo{author}{\bibfnamefont{R.}~\bibnamefont{Fern}},
  \bibinfo{author}{\bibfnamefont{R.}~\bibnamefont{Bondesan}}, \bibnamefont{and}
  \bibinfo{author}{\bibfnamefont{S.~H.} \bibnamefont{Simon}},
  \bibinfo{journal}{Phys. Rev. B} \textbf{\bibinfo{volume}{98}},
  \bibinfo{pages}{155321} (\bibinfo{year}{2018}),
  \urlprefix\url{https://link.aps.org/doi/10.1103/PhysRevB.98.155321}.

\bibitem[{\citenamefont{Zhu et~al.}(2018)\citenamefont{Zhu, Sheng, Fu, and
  Sodemann}}]{Zhu2018}
\bibinfo{author}{\bibfnamefont{Z.}~\bibnamefont{Zhu}},
  \bibinfo{author}{\bibfnamefont{D.~N.} \bibnamefont{Sheng}},
  \bibinfo{author}{\bibfnamefont{L.}~\bibnamefont{Fu}}, \bibnamefont{and}
  \bibinfo{author}{\bibfnamefont{I.}~\bibnamefont{Sodemann}},
  \bibinfo{journal}{Phys. Rev. B} \textbf{\bibinfo{volume}{98}},
  \bibinfo{pages}{155104} (\bibinfo{year}{2018}),
  \urlprefix\url{https://link.aps.org/doi/10.1103/PhysRevB.98.155104}.

\end{thebibliography}
\end{document}